\documentclass[aps,prd,notitlepage,showpacs,nofootinbib,superscriptaddress]{revtex4-1}
\pdfoutput=1
\usepackage[usenames,dvipsnames]{color}
\usepackage{graphicx}
\usepackage{setspace}
\usepackage{caption}
\usepackage{amsmath}
\usepackage{amssymb}
\usepackage[colorlinks=true,citecolor=darkred,urlcolor=darkred, pdfborder={0 0 0}]{hyperref}
\usepackage[normalem]{ulem}
\usepackage{xcolor}
\usepackage{color}
\usepackage{multirow}
\usepackage{tikz}

\makeatletter
\def\p@subsection{}
\makeatother
\definecolor{darkred}{rgb}{0.6,0,0}

\definecolor{linkcolor}{rgb}{0,0,0.5}

\def\gsim{\raise0.3ex\hbox{$\;>$\kern-0.75em\raise-1.1ex\hbox{$\sim\;$}}}
\def\lsim{\raise0.3ex\hbox{$\;<$\kern-0.75em\raise-1.1ex\hbox{$\sim\;$}}}

\def\beqn#1{\begin{equation}\label{#1}}
\def\eeqn{\end{equation}}

\def\beqa#1{\begin{eqnarray}\label{#1}}
\def\eeqa{\end{eqnarray}}

\usepackage{array,booktabs, makecell}
\usepackage{ulem} 
\usepackage{nicefrac}
\usepackage{tablefootnote}

\def\znbb {neutrinoless double beta decay }

\def\Z2{$\mathcal{Z_2}$}

\def\vev#1{\left\langle #1\right\rangle}

\newcommand {\ignore}[1]{}

\newcommand{\sm}{{Standard Model }}
\def\lfv{lepton flavour violation }

\def\SM{$\mathrm{SU(3)_c \otimes SU(2)_L \otimes U(1)_Y}$ }

\def\321{$\mathrm{SU(3) \otimes SU(2) \otimes U(1)}$ }

\def\red{\color{red}{}}
\def\black{\color{black}{}}

\def\dsf{$\Delta m_{\Sigma F}^{}$ } 
\def\depf{$\Delta m_{\eta^+ F}^{}$ } 
\def\deip{$\Delta m^2_{\eta_I^0\eta^+}$ }
\def\mdm{$m_{\text{DM}}^{}$ }
\allowdisplaybreaks

\begin{document}

\title{\boldmath \color{BrickRed} Phenomenological profile of scotogenic fermionic dark matter}

\author{Anirban Karan}\email{kanirban@ific.uv.es}
\affiliation{Institut de F\'{i}sica Corpuscular
, CSIC/Universitat de Valencia, Parc Cientific de Paterna, C/ Catedratico Jose Beltran, 2, E-46980 Paterna, (Valencia), Spain}
\author{Soumya Sadhukhan}\email{soumya.sadhukhan@rkmrc.in}
\affiliation{Ramakrishna Mission Residential College (Autonomous), Vivekananda Centre for Research, Narendrapur, Kolkata, India-700103} 
\affiliation{Institut de F\'{i}sica Corpuscular
, CSIC/Universitat de Valencia, Parc Cientific de Paterna, C/ Catedratico Jose Beltran, 2, E-46980 Paterna, (Valencia), Spain}
%\affiliation{\AddrAHEP}
\author{Jos\'{e} W. F. Valle}\email{valle@ific.uv.es}
\affiliation{Institut de F\'{i}sica Corpuscular
, CSIC/Universitat de Valencia, Parc Cientific de Paterna, C/ Catedratico Jose Beltran, 2, E-46980 Paterna, (Valencia), Spain}
%\affiliation{\AddrAHEP}

\begin{abstract}
  \vspace{0.3cm}

We consider the possibility that neutrino masses arise from the exchange of dark matter states.
We examine in detail the phenomenology of fermionic dark matter in the singlet-triplet scotogenic model. 
We explore the case of singlet-like fermionic dark matter, taking into account all co-annihilation effects relevant for determining its relic abundance,
such as fermion-fermion and scalar-fermion co-annihilation.
Although this in principle allows for dark matter below 60 GeV, the latter is in conflict with charged \lfv (cLFV) and/or collider physics constraints.
We examine the prospects for direct dark matter detection in upcoming experiments up to 10~TeV.
Fermion-scalar coannihilation is needed to obtain viable fermionic dark matter in the 60-100 GeV mass range.
 Fermion-fermion and fermion-scalar coannihilation play complementary roles in different parameter regions above 100~GeV.
   
\end{abstract}
%%%%%%%%%%%%%%%%%%%%%%%%%%%%
\maketitle 
%%%%%%%%%%%%%%%%%%%%%%%%%%%%
\vspace{-0.5cm} 
\section{Introduction} 
\label{sec:introduction}
\vspace{-0.3cm} 

The existence of neutrino masses~\cite{deSalas:2020pgw} and cosmological dark matter~\cite{Jungman:1995df,Bertone:2016nfn,Arcadi:2017kky} constitute two of the main pillars
indicating the need for new physics.
While they can be associated to totally unrelated sectors, the idea that dark matter mediates neutrino mass generation proposed independently by Ma~\cite{Ma:2006km}
and by Tao~\cite{Tao:1996vb} has by now become a paradigm
\cite{Hirsch:2013ola,Merle:2016scw,Rocha-Moran:2016enp,Restrepo:2019ilz,Choubey:2017yyn,Diaz:2016udz,Avila:2019hhv} for neutrino mass generation and for explaining cold dark matter in terms of weakly interacting massive particles (WIMPs).
     The main idea of these \textit{scotogenic} models~\cite{Tao:1996vb,Ma:2006km,Hirsch:2013ola,Merle:2016scw,Rocha-Moran:2016enp,Restrepo:2019ilz,Choubey:2017yyn,Diaz:2016udz,Avila:2019hhv}
     is to have neutrino masses generated radiatively (in our case at the one loop-level), mediated by a TeV-scale sector that can also account for WIMP dark matter (bosonic or fermionic).  
A $\mathcal Z_2$ symmetry is imposed in order to stabilize dark matter, ensuring also the radiative origin of neutrino masses.
By making one parameter small (namely $\lambda_5$) the symmetry of the model is enhanced, making it natural in the sense of 't Hooft \cite{tHooft:1979rat}.
In other words, neutrino masses are symmetry-protected.
Together with the loop-suppression in Fig.~\ref{fig:nu_mas} this protection makes it possible to keep the neutrino masses low in the presence of sizeable Yukawa couplings $\mathcal O (10^{-1})$
for the new fermion mediators. This makes the \textit{scotogenic} scenario phenomenologically very attractive in comparison with vanilla-type seesaw schemes.
Indeed, the simplest type-I seesaw scenarios based on the Standard Model~\cite{Schechter:1980gr,Schechter:1981cv} require either the neutrino
mass mediators to be superheavy or the Yukawa couplings too small, in order to fit the small observed neutrino masses. Hence they do not lead to observable phenomena,
e.g. in colliders. Finally, they require some ``external'' mechanism to account for dark matter, making these unrelated phenomena.

In this paper we focus on the singlet-triplet scotogenic model, proposed in \cite{Hirsch:2013ola}.
This offers a theoretically consistent picture of scotogenic dark matter in which the underlying $\mathcal Z_2$ symmetry
can be preserved up to high energies after renormalization group evolution of the couplings in the scalar sector~\cite{Merle:2016scw}.
Moreover, it offers a very rich, yet manageable phenomenology of direct dark matter detection, charged \lfv (cLFV) as well as collider physics.

Since the possibility of scalar scotogenic dark matter has already been considered~\cite{Diaz:2016udz,Avila:2019hhv},
  here we focus on the detailed study of fermionic scotogenic dark matter and its phenomenological features.
The latter is more intricately connected to the neutrino mass generation than its scalar dark matter counterpart,
where Higgs-portal-mediated annihilation processes play the dominant role, irrespective of the Yukawa couplings determining neutrino mass generation.
In contrast, fermionic dark matter is closely related with neutrino mass generation, and also involves the mixing of singlet and triplet neutral fermions. 
We note that singlet-like dark matter has a richer phenomenological profile than that of the pure triplet dark matter,
as the latter requires masses $\sim 3$~TeV in order to have enough co-annihilation to satisfy the relic density requirement.
We explore in detail the singlet-like fermionic dark matter, updating the experimentatal status and substantially improving upon
the early work in~\cite{Hirsch:2013ola} through a refined discussion of the relic abundance, in which we stress the important role of scalar-fermion
dark matter co-annihilation effects, which are properly taken into account.
Other novel features of our present work include a discussion of direct fermionic dark matter detection constraints,
combining with the restrictions of cLFV and collider experiments on dark matter phenomenology.
We explore the general feasibility of scotogenic fermionic dark matter, examining the prospects for direct dark matter detection
	in the next round of experiments all the way up to 10~TeV. 
We find that the possibility of light dark matter below 60 GeV or so is inconsistent with experimental cLFV and/or collider constraints.
Fermion-scalar coannihilation is required in order to obtain viable fermionic dark matter within the mass range of 60 GeV to 100 GeV.
  Beyond 100 GeV, fermion-scalar and fermion-fermion coannihilation play complementary roles at different regions of parameter space.

%%%%%%%%%%%%%%%%%%%%%%%%%%%%
\section{Singlet Triplet scotogenic Model} 
\label{sec:model}
%%%%%%%%%%%%%%%%%%%%%%%%%%%%%%%%

\vspace*{-3mm}

In this section we discuss the basic features of the singlet-triplet scotogenic model. 
Apart from the \sm (SM) particles, this model contains four colour-singlet states, two of which are fermions ($\Sigma$, $F$), the others are scalars ($\eta$, $\Omega$). 
The new fermionic states $\Sigma$ and $F$ are $SU(2)_L$ triplet and singlet respectively, with no hypercharge,
while the extra scalars $\eta$ and $\Omega$ are doublet and real triplet under the $SU(2)_L$ gauge group, with hypercharges $1/2$ and zero respectively.
In addition to the \SM gauge symmetry, a discrete $\mathcal Z_2$ symmetry is also imposed on the Lagrangian in order to ensure the stability of the lightest $\mathcal Z_2$-odd particle and its role as a suitable dark matter candidate.
Under this $\mathcal Z_2$ symmetry, all the SM particles along with the triplet scalar $\Omega$ are even, whereas the remaning new states are taken to be odd.   
The particle assignments (excluding the quarks) are given in \autoref{tab:part}, where $L$, $e$ and $\Phi$ denote the usual SM states corresponding to the left-handed lepton doublets,
right-handed charged lepton singlets and the scalar doublet. The hypercharge is assigned following the convention: $Q=T_3+Y$.
\renewcommand{\arraystretch}{1.2}
\begin{table}[h!]
	\centering
	\begin{tabular}{|c||>{\centering}m{0.7cm}|>{\centering}m{0.7cm}|>{\centering}m{0.7cm}||>{\centering}m{1cm}|>{\centering}m{1cm}||>{\centering}m{0.8cm}|>{\centering}m{0.8cm}|}
		\hline
		&\multicolumn{3}{c||}{Standard Model}&\multicolumn{2}{c||}{New Scalars}&\multicolumn{2}{c|}{New Fermions}\\
		\cline{2-8}
		&$L$&$e^c$&$\Phi$&$\Omega$&$\eta$&$\Sigma$&\multicolumn{1}{c|}{$F$}\\
		\hline\hline
		multiplicity &3&3&1&1&1&1&\multicolumn{1}{c|}{1}\\
		$U(1)_Y$ & $\nicefrac{-1}{2}$ & $1$& $\nicefrac{1}{2}$&0&$\nicefrac{1}{2}$&$0$&\multicolumn{1}{c|}{0}\\
		$SU(2)_L$ & 2 & 1 & 2 & 3 & 2 & 3 & \multicolumn{1}{c|}{1}\\
		$\mathcal Z_2$&+&+&+&$+$&$-$&$-$&\multicolumn{1}{c|}{$-$}\\
		\hline
	\end{tabular}
	\caption{
          Quantum number assignemnts for the singlet-triplet scotogenic model. The quark sector is standard.}
	\label{tab:part}
\end{table}

%%\vspace*{-5mm}

The relevant part of the Lagrangian involving the leptonic interactions for this model is written as:
%%%%%%%%%%%%%%%%%%%%%%
\begin{equation}
\label{eq:Yukawa}
- \mathcal{L}_{\rm Yuk} = -Y^{\alpha\beta} \bar L_{\alpha} \Phi e_{\beta} - Y_F^{\alpha} \bar L_{\alpha} \tilde{\eta} F^c - Y_{\Sigma}^{\alpha} \bar L_{\alpha} \Sigma^c \tilde{\eta} - Y_{\Omega} Tr[\bar{\Sigma}\Omega] F^c - \frac{M_{\Sigma}}{2} Tr[\bar{\Sigma}\Sigma^c] -  \frac{M_{F}}{2} \bar{F} F^c + \text{h.c.}
\end{equation}
%%%%%%%%%%%%%%%
where $Y^{\alpha\beta}$ are the usual Yukawa couplings for the SM leptons and $Y_F^{\alpha}, Y_{\Sigma}^{\alpha}, Y_{\Omega} $ are the Yukawa couplings involving the new fermions and scalars of the model.
Here the $\alpha,\,\beta$ are SM lepton family indices, while $M_{F}$ and $M_{\Sigma}$ denote Majorana mass terms for the additional fermions $F$ and $\Sigma$, respectively.
In the adjoint representation of $SU(2)_L$ the fermionic triplet $\Sigma$ and its charge conjugate $\Sigma^c$ are expressed as: 
\begin{equation}
\Sigma\equiv\begin{pmatrix}
\frac{\Sigma^0}{\sqrt 2} & \Sigma^+\\
\Sigma^- & -\frac{\Sigma^0}{\sqrt 2}
\end{pmatrix}\quad \text{and}\quad\Sigma^c\equiv\begin{pmatrix}
\frac{(\Sigma^0)^c}{\sqrt 2} & (\Sigma^-)^c\\
(\Sigma^+)^c & -\frac{(\Sigma^0)^c}{\sqrt 2}
\end{pmatrix}~,
\end{equation}
where the charged components form Dirac fermions.

%%%%%%%%%%%%%%%%%%%%%%%%%%%%%%%%%%%%%%%%%%%
\subsection{The scalar sector}  
\label{subsec:scalar-sector}
\vspace*{-3mm}
%%%%%%%%%%%%%%%%%%%%%%%%%%%%%%%%%%%%%%%%%%%
The scalar potential of this model, invariant under $SU(3)_C\otimes SU(2)_L\otimes U(1)_Y\otimes\mathcal{Z}_2$ symmetry, is given by:
\begin{align}
\label{eq:pot}
\mathcal{V}=&-\mu_\phi^2\,(\Phi^\dagger \Phi)+\mu_\eta^2 \,(\eta^\dagger \eta) -\frac{1}{2} \mu_\Omega^2\,\text{Tr}\,(\Omega^\dagger \Omega)
+\frac{1}{2}\lambda_1 (\Phi^\dagger\Phi)^2+\frac{1}{2}\lambda_2 (\eta^\dagger\eta)^2+\lambda_3 (\Phi^\dagger\Phi)(\eta^\dagger\eta)+\lambda_4 (\Phi^\dagger\eta)(\eta^\dagger\Phi)\nonumber\\
&+\frac{1}{2}\lambda_5 \Big[(\Phi^\dagger\eta)^2+(\eta^\dagger\Phi)^2\Big]+\frac{1}{2} \lambda^\phi_\Omega\, (\Phi^\dagger \Phi)\,\text{Tr}\,(\Omega^\dagger \Omega)+\frac{1}{4} \lambda_\Omega^\Omega\, [\text{Tr}\,(\Omega^\dagger \Omega)]^2+\frac{1}{2} \lambda^\eta_\Omega\, (\eta^\dagger \eta)\,\text{Tr}\,(\Omega^\dagger \Omega)\nonumber\\&+\mu^\phi_\Omega\,(\Phi^\dagger \Omega \Phi)+ \mu^\eta_\Omega\,(\eta^\dagger \Omega \eta)~,
\end{align}
where $\mu$'s are mass parameters and $\lambda$'s are dimensionless quartic couplings amongst different scalar fields.

After electroweak symmetry breaking (EWSB), the neutral components of $\Phi$ and $\Omega$ acquire non-zero vacuum expectation values (VEVs),
i.e. $v_\phi/\sqrt 2$ and $v_\Omega$, such that $\sqrt{v_\phi^2+v_\Omega^2}=v\approx246.22$ GeV.  
In contrast, the neutral component of the ``dark'' scalar $\eta$ does not get any VEV due to the conservation of $\mathcal Z_2$ symmetry. 
 Note that the triplet self-quartic interactions in general contain two contractions, $[\text{Tr}\,(\Delta^\dagger \Delta)]^2$ and $\text{Tr}\,[(\Delta^\dagger \Delta)^2]$.
However, since in our case the scalar triplet $\Omega$ is real and hyperchargeless, these terms are equivalent: $\text{Tr}\,[(\Omega^\dagger \Omega)^2]=\frac{1}{2} [\text{Tr}\,(\Omega^\dagger \Omega)]^2$. 
  Therefore it suffices to keep just one of them. 
Note also that $\text{Tr}[\Omega^\dagger\sigma_i \Omega]$ vanishes in our case for the same reason, which also eliminates quartic terms from the Higgs potential.

The scalar fields in the singlet-triplet scotogenic model can be expressed after EWSB as follows: 
\begin{align}
\Phi=
\frac{1}{\sqrt{2}}\begin{pmatrix}
\sqrt 2\,\phi^{+} \\
v_\phi+ \phi^0 +i A_\phi^0\\
\end{pmatrix},\quad
\Omega=
\begin{pmatrix}
\frac{v_\Omega}{\sqrt{2}}+\frac{\Omega^0}{\sqrt{2}} &   \Omega^{+} \\
\Omega^{-}  & -\frac{v_\Omega}{\sqrt{2}}-\frac{\Omega^0}{\sqrt{2}}\\
\end{pmatrix},\quad  \eta=
\frac{1}{\sqrt{2}}\begin{pmatrix}
\sqrt 2\,\eta^{+} \\
\eta_R^{0} +i\,\eta_I^{0}\\
\end{pmatrix}.
\end{align} 
Here $\phi^\pm$, $\Omega^\pm$ and $\eta^\pm$ are the charged partners of
the neutral CP-even states $\phi^0$, $\Omega^0$ and $\eta_R^0$, and neutral CP-odd states $A^0_\phi$ and $\eta_I^0$. 
Note that, in contrast to $\Phi$ and $\eta$, the neutral component of $\Omega$ has no pseudoscalar (CP-odd) particle, since $\Omega$ is a real triplet. 
Moreover, the charged components ($\Omega^\pm$) obey the relation $(\Omega^+)^*=\Omega^-$.
On the other hand, the pseudoscalar component $A^0_\phi$ is the Goldstone boson eaten up by the $Z$-boson.
Minimizing the scalar potential in  Eq. \eqref{eq:pot}, one gets two tadpole equations which allow us to express $\mu_{\phi}^2$ and $\mu_\Omega^2$ in the following fashion:
\begin{align}
\label{eq:tad}
\mu_\phi^2=\frac{1}{2}\lambda_1v_\phi^2+\frac{1}{2}\lambda_\Omega^\phi\, v_\Omega^2-\frac{1}{\sqrt 2}\, \mu_\Omega^\phi\, v_\Omega\,\quad\text{and} \quad
\mu_\Omega^2 =\lambda_\Omega^\Omega\, v_\Omega^2+\frac{1}{2}\lambda_\Omega^\phi\, v_\phi^2  -\frac{1}{2\sqrt 2}\,\mu_\Omega^\phi\, \frac{v_\phi^2}{v_\Omega}\,.
\end{align}
 
The squared mass matrix for the CP-even neutral components of $\mathcal Z_2$-even scalars, i.e. $\phi^0$ and $\Omega^0$, is given by:
\begin{equation}
\label{Eq:ScalarMM}
\mathcal{M}_{\phi^0}^2=\begin{pmatrix}
\lambda_1 v_\phi^2&& \lambda_\Omega^\phi v_\Omega v_\phi-\frac{1}{\sqrt 2}\, \mu_\Omega^\phi\, v_\phi\\
\lambda_\Omega^\phi v_\Omega v_\phi-\frac{1}{\sqrt 2}\, \mu_\Omega^\phi\, v_\phi&&2\lambda_\Omega^\Omega v_\Omega^2+\frac{1}{2\sqrt 2}\,\mu_\Omega^\phi\, \frac{v_\phi^2}{v_\Omega}
\end{pmatrix}.
\end{equation}
Diagonalization of this matrix leads to mixing of the scalars $\phi^0$ and $\Omega^0$ yield the mass eigenstates $h^0$ and $H^0$, one of which is the SM Higgs boson.
In our analysis, we have identified the lighter CP-even state $h^0$ as the SM Higgs boson with mass of 125.5 GeV. 
Likewise, the squared mass matrix for the $\mathcal Z_2$-even charged scalars $\phi^\pm$ and $\Omega^\pm$ take the form:
\begin{equation}
\label{eq:Hmass}
\mathcal{M}_{\phi^\pm}^2=\sqrt 2\, \mu_\Omega^\phi\,\begin{pmatrix}
 v_\Omega&& \frac{1}{ 2}\, v_\phi\\
\frac{1}{ 2}\, v_\phi&&\frac{1}{4}\, \frac{v_\phi^2}{v_\Omega}
\end{pmatrix}
+\;g^2\,\xi_{W^\pm} \begin{pmatrix}
\frac{1}{4} v_\phi^2 && -\frac{1}{2} v_\phi v_\Omega\\
-\frac{1}{2} v_\phi v_\Omega &&  v_\Omega^2
\end{pmatrix}
\end{equation}
where, $g$ is the SU(2) gauge coupling and $\xi_{W^\pm}$ is the parameter related to the gauge fixing term of $W^\pm$ boson.  
The charged bosons $\phi^{\pm}$ and $\Omega^{\pm}$ mix with each other through the diagonalization of $\mathcal{M}_{\phi^\pm}^2 $ matrix to provide the charged-scalar mass eigenstates,
one of which gets absorbed as the longitudinal $W^\pm$ Goldstone boson, while the second is a physical charged scalar state $H^{\pm}$. 
Note that although the mass of Goldstone boson, $m_{G^\pm}=\frac{1}{4}g^2\xi_{W^\pm}(v_\phi^2+4v_\Omega^2)$, depends on the choice of the gauge fixing parameter $\xi_{W^\pm}$,
  the mass of the physical charged scalar does not, $$m_{H^{\pm}}=\frac{\mu^\phi_\Omega}{2\sqrt 2\, v_\Omega}(v_\phi^2+4v_\Omega^2).$$

After EWSB, the squared masses for the $\mathcal Z_2$-odd or ``dark'' scalars and pseudoscalars are given by: 
\begin{eqnarray}
\label{eq:etmass}
&m_{\eta_R^0}^2=\mu_\eta^2+\frac{1}{2}\lambda_\Omega^\eta v_\Omega^2-\frac{1}{\sqrt 2} \mu_\Omega^\eta\, v_\Omega +\frac{1}{2}(\lambda_3+\lambda_4+\lambda_5)\,v_\phi^2\,, \qquad
m_{\eta_I^0}^2=\mu_\eta^2+\frac{1}{2}\lambda_\Omega^\eta v_\Omega^2-\frac{1}{\sqrt 2} \mu_\Omega^\eta\, v_\Omega +\frac{1}{2}(\lambda_3+\lambda_4-\lambda_5)\,v_\phi^2&\,,\nonumber\\
&m_{\eta^\pm}^2=\mu_\eta^2+\frac{1}{2}\lambda_\Omega^\eta v_\Omega^2+\frac{1}{\sqrt 2} \mu_\Omega^\eta\, v_\Omega +\frac{1}{2}\lambda_3\,v_\phi^2\,. &
\end{eqnarray}
Note that the squared masses of $\eta_R^0$ and $\eta_I^0$ differ by the term $\lambda_5 v_\phi^2$,
signaling the conservation of lepton number symmetry in the limit of vanishing $\lambda_5$.

\vspace*{-5mm}

\subsection{Neutrino mass generation}
\label{subsec:neutrino-mass}
\vspace*{-3mm}

From the Lagrangian in Eq.~(\ref{eq:Yukawa}) one sees that, once $\Omega^0$ acquires a VEV, the neutral component of the dark fermionic triplet, i.e. $\Sigma^0$,
mixes with dark singlet fermion $F$ through the term $Y_{\Omega} Tr[\bar{\Sigma}\Omega] F^c$. 
The corresponding mass matrix representing the mixing of F and $\Sigma_0$ is given by: 
\begin{equation}
\mathcal M_\chi =\begin{pmatrix}
M_F & Y_\Omega v_\Omega\\
Y_\Omega v_\Omega & M_\Sigma
\end{pmatrix}.
\end{equation}
Due to the Pauli principle, the matrix $\mathcal M_\chi$ is in general symmetric, but complex, and can always be diagonalized by a unitary matrix $V$ 
involving an angle $\theta$ and one Majorana phase~\cite{Schechter:1980gr}. 
Here for simplicity we assume CP conservation in the dark fermion sector, so that $\mathcal M_\chi$ is diagonalized by an orthonormal transformation, as follows:
\begin{eqnarray}
\begin{pmatrix}
\chi_1^{0} \\ \chi_2^{0}
\end{pmatrix}= V \cdot \begin{pmatrix}
F\\ \Sigma^0
\end{pmatrix} \quad 
\text{with} \quad
V=\begin{pmatrix}
\cos\theta & -\sin\theta\\
\sin\theta & \cos \theta
\end{pmatrix},
\quad
\text{such that} \quad V\cdot\mathcal M_\chi\cdot V^T=\text{diag}(m_{\chi_{1}^{0}},m_{\chi_{2}^{0}}),
\end{eqnarray}
where the mixing angle $\theta$ and the masses of physical states $m_{\chi_{1,2}^{0}}$ can be expressed as:
\begin{equation}
\label{eq:tht_mch}
\tan(2\theta)=\frac{2\,Y_\Omega\, v_\Omega}{M_\Sigma-M_F}, \quad m_{\chi_{1,2}^{0}}=\frac{1}{2}\Big[(M_\Sigma+M_F)\mp\sqrt{(M_\Sigma-M_F)^2+ 4Y_\Omega^2\,v_\Omega^2}\Big] .
\end{equation}

%%%%%%%%%%%%%%%%%%%%%%%%%%%%%%%%%%%%%%%%%%%%%

\begin{figure}[h!]
	\includegraphics[scale=0.25]{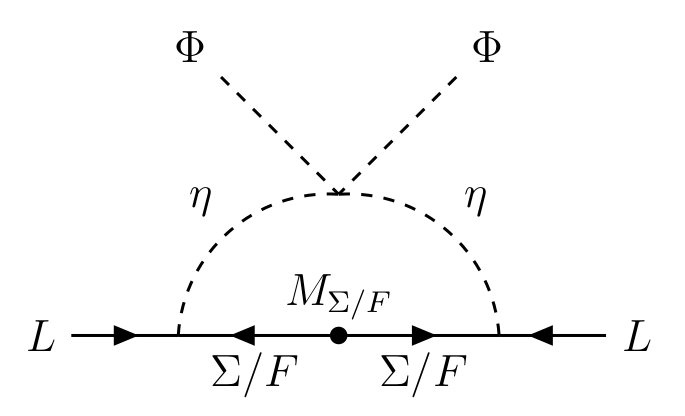}\hfil\raisebox{-8.0mm}[0pt][0pt]{\includegraphics[scale=0.25]{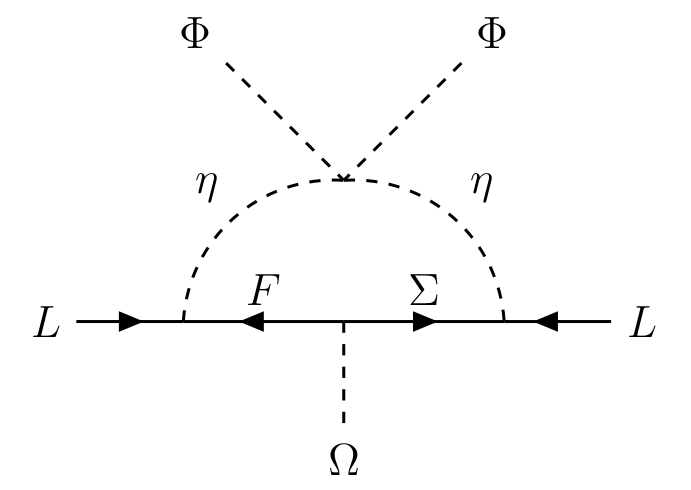}}
	\vspace*{5mm}
	\caption{Feynman diagrams for neutrino mass generation in one loop. }
	\label{fig:nu_mas}
      \end{figure}

      The lightest of these two dark Majorana fermions $\chi_j^0$ is satibilized by the $\mathcal Z_2$ symmetry, serving as our fermionic dark matter candidate. 
      The mixing angle $\theta$ determines the amount of singlet or triplet composition of our WIMP dark matter candidate.
  These states are analogous to the Bino and neutral Wino dark matter candidates in R-parity conserving supersymmetic models. % \\[-3cm]     

  Following the scotogenic paradigm, the neutrino mass-matrix is generated at one-loop level, as shown in Fig. \ref{fig:nu_mas}.
  Majorana masses for neutrinos are mediated by dark sector exchange, and extracted from: 
  \begin{align}
&\mathcal M_\nu=Y_\nu\cdot\mathcal F\cdot Y_\nu^T,\quad
\text{with}\quad  Y_\nu=\begin{pmatrix}
Y_F^1&\frac{Y_\Sigma^1}{\sqrt 2}\\
Y_F^2&\frac{Y_\Sigma^2}{\sqrt 2}\\
Y_F^3&\frac{Y_\Sigma^3}{\sqrt 2}
\end{pmatrix}\cdot V^T(\theta)\quad \text{and} \quad   \mathcal F=\begin{pmatrix}
\frac{\mathcal I_1}{32\pi^2}&0\\
0&\frac{\mathcal I_2}{32\pi^2}
\end{pmatrix},\nonumber\\
& \quad\text{where,} \quad \mathcal I_j=m_{\chi_j^{0}}\Bigg[\frac{\ln\Big(m_{\chi_j^{0}}^2/m_{\eta_R^{0}}^2\Big)}{\Big(m_{\chi_j^{0}}^2/m_{\eta_R^{0}}^2\Big)-1}-\frac{\ln\Big(m_{\chi_j^{0}}^2/m_{\eta_I^{0}}^2\Big)}{\Big(m_{\chi_j^{0}}^2/m_{\eta_I^{0}}^2\Big)-1}\Bigg]\quad\text{for} \quad j\in \{1,2\}.
  \end{align}
  Note that in the limit of vanishing $\lambda_5$ lepton number symmetry would be restored, the states $\eta_{R,I}^0$ become degenerate,
  making the functions $\mathcal I_j$ and hence the neutrino mass matrix $\mathcal M_\nu$ to vanish. 
  Thus the choice of $\lambda_5\ll 1$ is natural in the sense of 't Hooft \cite{tHooft:1979rat} as the limit of $\lambda_5\to0$ enhances the  symmetry of the model.\\[-.2cm]
      
  In the singlet-triplet scotogenic model the lightest neutrino is massless, as in all missing partner (radiative) seesaw schemes~\cite{Schechter:1980gr,Ibarra:2003up}.
  In the same spirit as Ref.~\cite{Casas:2001sr}) one can extract the Yukawa matrix $Y_\nu$ in such a way as to automatically satisfy the neutrino oscillation data as: 
\begin{equation}
\label{eq:casas}
Y_\nu=U\cdot(\widetilde{\mathcal M}_\nu)^{1/2}\cdot\rho\cdot(\mathcal F)^{-1/2}\quad \text{with}\quad U^\dagger \mathcal M_\nu U^*=\widetilde{\mathcal M}_\nu~,
\end{equation}
where $\widetilde{\mathcal M}_\nu$ is the neutrino mass matrix in diagonal form and $U$ is the lepton mixing matrix, both measured to a large extent. 
On the other hand, the complex matrix $\rho$ can be paramatrized for normal and inverted neutrino mass-ordering through a complex angle $\omega$ as \cite{Ibarra:2003up}: 
\begin{equation}
\rho^{\rm NO}=\begin{pmatrix}
0&0\\
\cos\omega&\sin\omega\\
-\sin\omega& \cos\omega
\end{pmatrix}\quad \text { and }\quad 
\rho^{\rm IO}=\begin{pmatrix}
\cos\omega&\sin\omega\\
-\sin\omega& \cos\omega\\
0&0
\end{pmatrix}.
\end{equation}

\section{ Constraints}  
\label{sec:constr}

As mentioned above, it is useful to extract the Yukawa matrix $Y_\nu$ in terms of the measured neutrino oscillation data as in Eq.~(\ref{eq:casas}).
This will expedite the scan procedure we perform in order to determine the viable model parameter space.
Before doing this however, let us first compile the restrictions we have imposed on the numerical scan of the model parameters.

\subsection{Theoretical Constraints and Electroweak Precision Observables } 
\label{sub:pert} 

\begin{itemize}
	
\item {\bf Theoretical Constraints} \\[-3mm]

  In order to prevent the potential acquiring large negative values at large field values, the following conditions \cite{Kannike:2012pe,Merle:2016scw}
  should be obeyed by the quartic couplings of this model: 
	\begin{align}
		\label{eq:th_cons}
	&\hspace*{-0.2cm}i)\;\lambda_1\geq 0, \qquad ii)\;\lambda_2\geq 0, \qquad iii)\;\lambda_\Omega^\Omega\geq 0,\qquad iv)\;\lambda_3+\sqrt{\lambda_1 \lambda_2}\geq 0,\nonumber\\
	 &\hspace*{-0.2cm}v)\;\lambda_3+\lambda_4-|\lambda_5|+\sqrt{\lambda_1 \lambda_2}\geq 0,\qquad vi)\;\lambda_\Omega^\phi+\sqrt{2\lambda_1 \lambda_\Omega^\Omega} \geq 0, \qquad vii)\;\lambda^\eta_\Omega+\sqrt{2\lambda_2 \lambda_\Omega^\Omega} \geq 0,\nonumber\\
	&\hspace*{-0.2cm}viii)\;\sqrt {2\lambda_1\lambda_2\lambda_\Omega^\Omega}+\lambda_3\sqrt{2\lambda_\Omega^\Omega}+\lambda_\Omega^\phi\sqrt\lambda_2+\lambda^\eta_\Omega\sqrt \lambda_1+\sqrt{\Big(\lambda_3+\sqrt{\lambda_1 \lambda_2}\Big)\Big(\lambda_\Omega^\phi+\sqrt{2\lambda_1\lambda_\Omega^\Omega}\Big)\Big(\lambda^\eta_\Omega+\sqrt{2\lambda_2\lambda_\Omega^\Omega}\Big)}\geq 0,
		\end{align}
                when $\lambda_4+|\lambda_5|\geq 0$.
                On the other hand, if $\lambda_4+|\lambda_5|< 0$ holds, the last inequality (viii) is modified by replacing the $\lambda_3$ term by
                $(\lambda_3+\lambda_4-|\lambda_5|)$. 
                This way we can keep the scalar potential  Eq. \eqref{eq:pot} bounded from below.

              Moreover, to ensure perturbativity, the absolute value for all the dimensionless quartic couplings ($\lambda_i$) in the scalar potential should be smaller
              than $4\pi$, while the Yukawa couplings ($Y^{\alpha\beta}$, $Y_F^\alpha$, $Y_\Sigma^\alpha$, $Y_\Omega$) should be smaller than $\sqrt{4\pi}$ at any particular energy scale.
%%           %
              Although the exact values of the Yukawa and quartic couplings' upper limits coming from perturbativity are model-dependent \cite{Allwicher:2021rtd, Bahl:2022lio},
              the above mentioned values are pretty common choices in the literature~\cite{Bandyopadhyay:2021kue}.  
              We also mention that the renormalization group running of different parameters of this model may lead to the breaking of $\mathcal Z_2$ symmetry \cite{Merle:2016scw}.
              This can be avoided by choosing $\mu_\Omega^\eta \lesssim \mathcal{O} (1 \text{ TeV})$.

%\subsection{ S, T, U parameters}
\item {\bf S, T, U parameters} \\[-3mm]
  
 It is well-known that the presence of Higgs triplet affects the $\rho$-parameter at tree-level~\cite{Schechter:1980gr}.
 Indeed, our triplet $\Omega$ contributes to the mass of $W$ boson (but not to the $Z$ boson mass), so that the $\rho$-parameter at tree-level becomes~\cite{Diaz-Cruz:2003kcx}  
	\begin{equation}
		\rho=\frac{\sum_{i}c_i\langle\phi_i^0\rangle^2[T_i(T_i+1)-Y_i^2]}{2\sum_{i}^{}\langle\phi_i^0\rangle^2 Y_i^2}\Bigg|_ {i\in\{\Phi,\Omega\}}=1+\frac{4\,v_\Omega^2}{v_\phi^2}
              \end{equation}
              Here $\vev{\phi_i^0}$ is the VEV of the neutral component of  $\phi_i$, $T_i$ and $Y_i$ are the weak isospin and the hypercharge of $\phi_i$, and $c_i$ is a constant that equals to 1/2 or 1
              depending on the scalar being in real or complex representations (note that Ref. \cite{Diaz-Cruz:2003kcx} uses the convention: $Q=T_3+Y/2$).
              
              Hence the precise measurement of the $\rho$-parameter puts a severe restriction on $v_\Omega$.
              The current global fit for $\rho$-parameter is $\rho=1.00038\pm 0.00020$ \cite{PDG}, nearly $2\sigma$ away from the tree-level SM expectation.  
              Taking the $3\sigma$ range of the fit, one can restrict the VEV of triplet as: $v_\Omega\lsim 4$ GeV. 
 This maximum value of $v_\Omega$ can push the mass of W-boson up to 80.389 GeV which is consistent with the PDG value \cite{PDG} at $1\sigma$. 
                            
              The current global fit yields the following S, T and U parameter values~\footnote{There are non-negligible correlations among them too.}:
      \begin{equation}
 	S=-0.02 \pm 0.10,\quad T=0.03\pm 0.12, \quad U= 0.01\pm 0.11,
 \end{equation}

which are in good agreement with SM prediction. However, the error in the fits leave lots of room for different new physics scenarios.
Indeed, the recent CDF W-mass measurement \cite{CDF:2022hxs} can weaken the above limits~\footnote{Interesting models and phenomenological discussions on the W-mass shift
 can de found in~\cite{Batra:2022arl,Kanemura:2022ahw,Biekotter:2022abc,VanLoi:2022eir,Batra:2022pej,Batra:2022org}.}. 

\subsection{Neutrino Oscillation Constraints} 
\label{sub:neu-osci}
\vspace*{-3mm}
 	
%\item {\bf Neutrino Oscillation Data} \\
 
          As already mentioned, in our singlet-triplet scotogenic model the lightest neutrino is massless, a characteristic feature of missing partner seesaw schemes
          containing an incomplete set (less than 3) of non-doublet neutrino species~\cite{Schechter:1980gr,Ibarra:2003up}.
          For definiteness here we take normal neutrino-mass-ordering, as preferred by the global oscillation fits~\cite{deSalas:2020pgw}, so that $m_1 = 0$. 
          The global fit of the neutrino oscillation related constraints are in the best fit point with $ \pm 1 \sigma$, taken from Reference~\cite{deSalas:2020pgw}: 
	\begin{eqnarray*}
		&\sin^2 \theta_{12} = 0.304^{+0.012}_{-0.012}, \ \ \sin^2 \theta_{23} = 0.537^{+0.016}_{-0.020}, \ \ \   \sin^2 \theta_{13} = 0.0022^{+0.00062}_{-0.00063},&\\	
		&\Delta m_{21}^2 = 7.42^{+0.20}_{-0.21} \times 10^{-5} \ \rm eV^2, \ \ \
		\Delta m_{31}^2 = 2.517^{+0.026}_{-0.028} \times 10^{-3} \ \rm eV^2, \ \ \
		\delta_{\rm CP} = {197^\circ}^{+27^\circ}_{-24^\circ}&.	
	\end{eqnarray*}
	We use the $3 \sigma$ range for these, i.e. $\sin^2 \theta_{12} : (0.269 \to 0.343)$, $\sin^2 \theta_{23}:(0.415 \to 0.616)$, $\sin^2 \theta_{13}:(0.02032 \to 0.02410).$
        The corresponding ranges for the mass differences are $\Delta m_{21}^2: (6.82 \to 8.04) \times 10^{-5} \ \rm eV^2$ and $\Delta m_{31}^2: (2.435 \to 2.598) \times 10^{-3} \ \rm eV^2$. 
	In our analysis we vary the values these randomly in different Gaussian distributions (with means taken as the experimentally measured values and standard deviations taken as the $1\sigma$ error bar)
        truncated at the $3\sigma$ level. Concerning the above global fit results we notice that they are in good agreement with those of the other groups~\cite{Esteban:2020cvm,Capozzi:2021fjo}.

\vspace*{-3mm}	

\subsection{Collider Constraints} 
\label{sub:Collider}
\vspace*{-4mm}

The new scalars as well as dark fermions present in the singlet-triplet scotogenic model are subject to restrictions arising from existing collider searches, which we now summarize.

\item {\bf Direct Search Constraints}\\  
  $\bullet$ Several searches have been performed to detect charged scalars at LEP and LHC.
  The LEP bound \cite{ALEPH:2013htx} on charged scalars ($m_{H^{\pm}} \ge 80 $~GeV) assumes that the charged scalar is part of a $SU(2)_L$ doublet (weak isospin 1/2) and that
  the branching fractions of charged scalar to $\tau^+\nu$ and  $c\bar s$ add up to unity \cite{PDG}. 
  On the other hand ATLAS \cite{ATLAS:2014otc} and CMS \cite{CMS:2015lsf} exclude charged scalars in the [80 – 140] GeV and [90 – 155] GeV ranges, respectively. 
  These searches place constraints on scenarios where the quarks directly couple to the new scalar, which is not true for our case regarding both ($\eta^\pm$ and $H^\pm$).
  Therefore, their production cross-section at the LHC will be very suppressed and these bounds cannot be applied strictly to our scenario.

However, the searches for supersymmetric particles, especially slepton-pair-production, followed by their decay to SM leptons and neutralinos,
at LEP \cite{OPAL:2003wxm,OPAL:2003nhx} and the LHC \cite{ATLAS:2019lff,ATLAS:2022hbt} can be used to constrain the masses of $\eta^\pm$ and
$H^\pm$. 
The LHC result for slepton decaying to lepton and massless neutralino rules out $H^\pm$ below the mass of 400 GeV (considering a conservative limit \cite{ATLAS:2019lff,ATLAS:2022hbt}).
Again, the constraints from the LEP require $\eta^\pm$ to be heavier than 70 GeV \cite{Pierce:2007ut}.

$\bullet$ Additional neutral scalars have also been searched for at LEP and LHC, mainly in the context of 2-doublet Higgs models.
Using the di-boson ($\gamma\gamma$, $WW^*$ and $ZZ^*$) decay channels they exclude some mass range for the extra neutral scalars.  
The L3 collaboration provides the most severe bound on the lowest mass allowed for this fermiophobic scalar as 107 GeV~\cite{L3:2003ieq},
while the Tevatron reports the exclusion of any such scalar in the [100-116]~GeV mass range \cite{CDF:2013kiv}.  
Di-photon decay modes in ATLAS and CMS rule out such scalars in [110--121] \cite{ATLAS:2012yxc} and [110--147] GeV~\cite{CMS:2013zma}, respectively. 
On the other hand, assuming that the second Higgs decays invisibly, ALEPH puts the bound on the lowest allowed mass of any additional neutral scalar to be 114 GeV \cite{ALEPH:2001roc}
using the single production channel of such scalar associated with a $Z$-boson. 
These constraints may be applicable to $H^0$ pushing its mass above 150 GeV or so.

Note however that these constraints are not relevant for the $\mathcal Z_2$-odd neutral scalars ($\eta_R^0$ and $\eta_I^0$).
Indeed, the $\mathcal Z_2$-symmetry prevents these to be singly produced and also to decay to di-boson channels.
Nevertheless, LEP searches \cite{OPAL:2003wxm,OPAL:2003nhx} for neutralinos constrain the masses of $\eta_R^0$ and $\eta_I^0$. 
For instance, using the LEP data in the case of inert two Higgs doublet model Ref. \cite{Lundstrom:2008ai} excludes the region where the
three conditions of $m_{\eta_R^0}<80$ GeV, $m_{\eta_I^0}<100$ GeV and $m_{\eta_I^0}-m_{\eta_R^0}>8$ GeV are satisfied simultaneously.

$\bullet$ LEP searches set a lower limit on the mass of heavy charged lepton as 102 GeV~\cite{OPAL:2003zpa,L3:2001xsz} assuming it to be long-lived or stable.
Therefore, we take this into account while considering mass of $\Sigma^+$, i.e. $M_\Sigma$.
For the heavy neutral fermions, LEP puts lower limit on fermion masses at 102 GeV and 90 GeV \cite{L3:2001xsz} for Dirac and Majorana fermions respectively,
assuming their decay to SM charged leptons plus the $W$-boson.
In our model there are two neutral leptons, $\chi_{1,2}^{0}$, where the lightest is the dark matter candidate and hence stable by the $\mathcal Z_2$-symmetry.  
Even the heavier $\mathcal Z_2$-odd neutral fermion cannot decay to two SM particles, so the LEP bound is not directly applicable. 
Moreover, since $m_{\chi_{2}^{0}}$ is always bigger than $M_\Sigma$, imposing the lower limit on $M_\Sigma$ as 102 GeV automatically sets the same limit for $m_{\chi_{2}^{0}}$.

It is interesting to mention that the LHC \cite{ATLAS:2020wop} rules out $\mathcal Z_2$-even heavy triplet fermion up to 790 GeV of mass. 
But this result does not directly affect our $\mathcal Z_2$-odd triplet $\Sigma$ since the final states considered at the LHC are forbidden in our model.

\black
\item {\bf Z/W Widths and Higgs Invisible Decay} \\
  SM gauge boson ($W/Z$) decays have been measured with a great precision. New light particles could show up in Z and W decays. 
  Whenever kinematically allowed, such additional decay channels will affect the widths of the Higgs, $Z$ and $W$ bosons.
  Likewise, Higgs decays involving dark channels contribute to the invisible width, and can also be constrained to some extent. 
  The expressions for the $Z$, $W$ and Higgs boson partial decay widths are listed in Appendix~\ref{sec:width}. 
  For our analysis, we have used following constraints: 
\begin{itemize}
\item the branching fraction of the Higgs boson to the invisible modes should be less than $13\%$~\cite{ATLAS:2022vkf}. 

\item new contributions to the total $Z$-boson width should be less than 5 MeV (i.e. roughly within $2\sigma$), as the error in the measurement is 2.3 MeV~\cite{PDG}.

\item new contributions to the total $W$-boson width should be less than 90 MeV (i.e. almost within $2\sigma$), as the error in the measurement is 42 MeV~\cite{PDG}.
\end{itemize}
Here we have adopted a conservative approach to accommodate new contributions within the $2\sigma$ bound of the experimentally measured values.
These uncertainties in the widths of W/Z help in restricting the BSM parameters of the model while keeping the $W$ mass fixed
at: $m_W^2=\frac{1}{4} g^2(v_\phi^2+4v_\Omega^2)$.

\end{itemize}

\black
\section{Parameter Simulation Procedure}

Besides the SM quark and lepton Yukawa couplings, the model contains many free parameters describing its Yukawa ( Eq. \eqref{eq:Yukawa}) as well as scalar ( Eq. \eqref{eq:pot}) sector.
They are tabulated in Table.~\ref{TabP}. 
\begin{table}[h!]
	\begin{center}
		\begin{tabular}{| c | c | c | c | c |}
			\hline 
			Complex Yukawa  & \,Real Yukawa\, & \,Scalar mass terms\,  & \,\multirow{2}{*}{Scalar quartic couplings}\, & \,\multirow{2}{*}{Fermionic mass terms}\,\\
			couplings\footnote{Each complex quantity counts as two parameters.}&couplings\footnote{To simplify our analysis we assume $Y_\Omega$ to be real.}&and trilinear couplings&&\\
			\hline \hline			 
			$Y_F^1$, $Y_F^2$, $Y_F^3$,  & \multirow{2}{*}{$Y_\Omega$} &  \multirow{2}{*}{$\mu_{\phi}$, $\mu_{\eta}$, $\mu_{\Omega}$, $\mu_{\Omega}^{\phi}$, $\mu_{\Omega}^{\eta}$} &   $\lambda_1$, $\lambda_2$ $\lambda_3$, $\lambda_4$, $\lambda_5$, &   \multirow{2}{*}{$M_F$, $M_\Sigma$} \\	
			$Y_\Sigma^1, Y_\Sigma^2, Y_\Sigma^3$  &  &   &  $\lambda_{\Omega}^{\phi}$, $\lambda_{\Omega}^{\eta}$, $\lambda_{\Omega}^{\Omega}$ &    \\
			\hline %\hline
		\end{tabular}
	\end{center}
	\caption{
          Free model parameters consist of five $\mu_i$ parameters, eight $\lambda_i$, three complex $Y_F^\alpha$, three complex, $Y_\Sigma^\alpha$, $Y_\Omega$,
          and two fermionic masses $M_F$ and $M_\Sigma$.}
	\label{TabP} 
      \end{table}
      
      However, not all of these parameters are independent, since some can be cast in terms of known quantities, thereby reducing the number of independent parameters. 
      For example the Higgs VEV and mass are determined, reducing two parameters, whereas neutrino oscillation data determine the two light neutrino masses (there are only two in our scheme),
      the three mixing angles and one of the CP phases, though rather poorly so far. 
      Thus the Yukawa couplings $Y_F$ and $Y_\Sigma$ can be traded only in terms of complex angle $\omega$ using  Eq. \eqref{eq:casas}.
      Neglecting the Majorana phase this results in sixteen independent parameters to deal with.

      Instead of working directly with the parameters in \autoref{TabP}, it is useful to use other parameters which can be expressed as combinations of the above.
      For example, instead of working with $\mu$'s, it is more helpful to work with physical scalar mass parameters and VEVs.  
      In fact, since we will be studying co-annihilation effects, the mass differences or squared differences are more useful parameters than the masses themselves. 
      Therefore we define the parameters $\Delta m_{\Sigma F}$\,, $\Delta m_{\eta^+ F}$ and $\Delta m^2_{\eta_I^0\eta^+}$ in the following way:
\begin{equation}
	\Delta m_{\Sigma F}^{}=M_\Sigma-M_F, \qquad \Delta m_{\eta^+ F}=m_{\eta^+}-M_F \qquad\text{and} \qquad \Delta m^2_{\eta_I^0\eta^+}=m^2_{\eta^0_I}-m^2_{\eta^+}~.
\end{equation}
  Two of the $\mu$'s can be removed in favor of the measured electroweak VEV and the SM Higgs mass (using  Eq. \eqref{eq:tad} and  Eq. \eqref{Eq:ScalarMM}),
  and two more of them can be traded in favor of $v_\Omega$ and \depf (with the help of  Eq. \eqref{Eq:ScalarMM} and  Eq. \eqref{eq:etmass}). This way we keep only one
  of the five $\mu$'s, say $\mu_\Omega^\eta$, as independent. 
Furthermore, $\lambda_4$ can be replaced easily by \deip using  Eq. \eqref{eq:etmass}.
Then the set of independent parameters to deal with become: $\mu_\Omega^\eta$, $v_\Omega$, $Y_\Omega$, seven $\lambda_i$ ($\lambda_4$ is not independent), four parameters related to
$\mathcal{Z}_2$-odd particle masses, i.e. $M_F$\,, $\Delta m_{\Sigma F}$\,, $\Delta m_{\eta^+ F}$\,, $\Delta m^2_{\eta_I^0\eta^+}$ and the complex angle $\omega$.
\renewcommand{\arraystretch}{1.5}
\begin{table}[h!]
	\scalebox{1.1}{
			\begin{tabular}
					{||c|c|c|c||c|c|c||c|c||c|c|c|c|c|c|c||}
					\hline
					$M_F$& $\Delta m_{\Sigma F}$& $\Delta m_{\eta^+ F}$& $\Delta m^2_{\eta_I^0\eta^+}$&$\mu_\Omega^\eta$& $v_\Omega$& \multirow{2}{*}{$Y_\Omega$} &\multirow{2}{*}{Re$(\omega)$}&\multirow{2}{*}{Im$(\omega)$}&\multirow{2}{*}{$\lambda_1$}&\multirow{2}{*}{$\lambda_2$}&\multirow{2}{*}{$\lambda_3$}&\multirow{2}{*}{$\lambda_5$}&\multirow{2}{*}{$\lambda_\Omega^\phi$}&\multirow{2}{*}{$\lambda_\Omega^\Omega$}&\multirow{2}{*}{$\lambda_\Omega^\eta$}\\
					(GeV)&(GeV)&(GeV)&(GeV$^2$)&(GeV)&(GeV)&&&&&&&&&&\\
					\hline
					[1, 1000]&200&500&1000&400&4.0&2.0&$\pi/4$&$\pi/4$& 0.2626 &0.5&0.5&$10^{-8}$&0.5&0.5&0.5\\
					\hline
			\end{tabular}}
	\caption{Specification of the benchmark scenario BP0.}
	\label{tab:BP0}
      \end{table}
      
Our fermionic scotogenic dark matter scans require a multi-parametric setup. For definiteness we choose a benchmark point BP0 specified by the parameter values given in \autoref{tab:BP0}.  
We change each parameter separately and study the effect on the various observables, implementing also the constraints discussed above.
With this guidance we proceed with the parameter scanning.

The values of $Y_\Omega$, $\lambda_i$ and $\omega$ in BP0 satisfy all the conditions required in Sec. \ref{sub:pert}. 
The chosen value of $\mu_\Omega^\eta$ preserves the $\mathcal Z_2$ symmetry at high scales, while particle masses satisfy the relevant collider constraints mentioned in Sec. \ref{sub:Collider}. 
We choose the second $\mathcal Z_2$-even scalar of our model ($H$) as heavier than the SM Higgs ($h$), resulting in the non-dark charged scalar ($H^+$) being heavier than $h^0$. 
One can also check that the benchmark point BP0 does not add new contributions to the Z/W width or Higgs invisible decay. 
Since the mass differences between charged and neutral components of $\Omega$ and $\eta$ are not too big, the T-parameter does not change much.
As the triplet scalar is heavy ($m_H\sim$ 400 GeV), the S-parameter remains also almost the same. 
Since U is the coefficient of the dimension-eight operator in the EFT approach, it usually remains very small for a wide class of BSM scenarios. 
All in all, BP0 is a valid benchmark point for study. \\[-.3cm] 

We first examine the prospects of cLFV and fermionic scotogenic dark matter using the benchmark point BP0. 
Our main goal with BP0 is to study the dependence of cLFV processes and dark matter relic density on the parameters of this model,
as this serve as guide towards the scanning of the parameter space.
To obtain our numerical results, the model has been implemented in {\tt SARAH} \cite{Staub:2008uz,Staub:2013tta,Staub:2015kfa}.
The cLFV prospects have been studied using the package {\tt SPheno} \cite{Porod:2003um,Porod:2011nf} that employs the {\tt FlavourKit} \cite{Porod:2014xia} component of {\tt SARAH}.
 On the other hand, the dark matter analyses have been performed using {\tt micrOMEGAs} \cite{Belanger:2001fz,Belanger:2020gnr}.

\section{Charged Lepton Flavor Violation}
\label{sec:lfv-results}

\begin{table}[h!]
	\begin{tabular}{| c | c | c |}
		\hline
		\;	LFV Process \; & \; Current Bound \; & \; Future Sensitivity \;\\
		\hline
		$\mathcal B (\mu\to e \gamma)$& $4.2 \times 10^{-13}$ \cite{MEG:2016leq} & $6.0 \times 10^{-14}$ \cite{MEGII:2018kmf}\\
		\hline
		$\mathcal B (\mu\to 3e)$ & $1.0 \times 10^{-12}$ \cite{SINDRUM:1987nra} & $\sim 10^{-16}$ \cite{Blondel:2013ia,Mu3e:2020gyw}\\
		\hline
		$\mathcal C (\mu, Au \to e, Au)$ & $7.0\times 10^{-13}$ \cite{SINDRUMII:2006dvw} & -- \\
		\hline
		$\mathcal C (\mu, Ti \to e, Ti)$ & $4.3\times 10^{-12}$ \cite{SINDRUMII:2006dvw} &  $\sim 10^{-18}$ \cite{prism}\\
		\hline
		$\mathcal C (\mu, Pb \to e, Pb)$ & $4.6\times 10^{-11}$ \cite{SINDRUMII:2006dvw} &  -- \\
		\hline
		$\mathcal C (\mu, Al \to e, Al)$ & -- & $\sim 10^{-17}$ \cite{Mu2e:2014fns,COMET:2018auw}\\
		\hline
	\end{tabular}
	\caption{The experimental bounds on cLFV processes involving $\mu$ decays or conversions.}
	\label{tab:LFV}	
\end{table} 

\begin{figure}[h!]
	\vspace*{3mm}
	\includegraphics[scale=0.22]{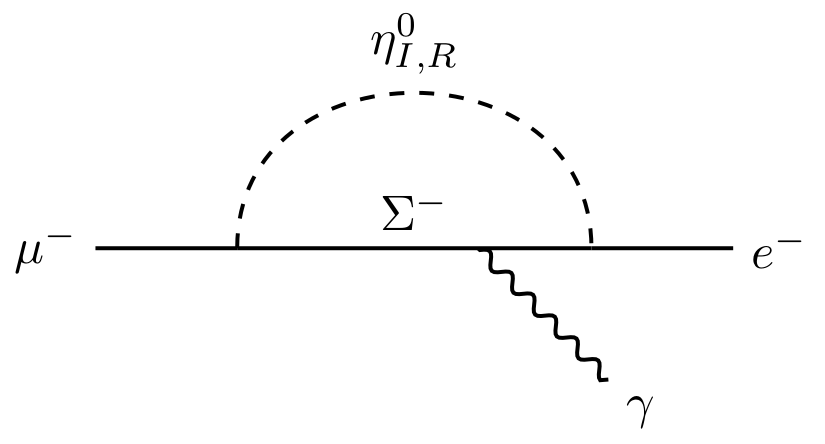}\hfil\raisebox{11.0mm}[0pt][0pt]{\includegraphics[scale=0.22]{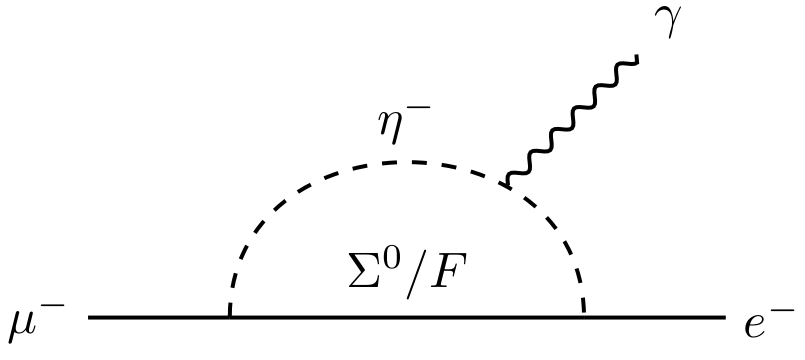}}
	\caption{Feynman diagrams for ``dark-mediated'' $\mu \to e \gamma$.}
	\label{fig:mutoe}
\end{figure}

The conventional charged current mechanism leading to charged lepton flavour violating processes such as $\mu\to e \gamma$ is highly suppressed due to the GIM mechanism.
However, the mediators responsible for neutrino mass generation also mediate rare processes with charged lepton flavour violation, such as $\mu\to e \gamma$.
Indeed, we expect these new Feynman diagrams (see Fig. \ref{fig:mutoe}) to lead to enhanced rates for these rare decays, as examined in detail in Ref.~\cite{Rocha-Moran:2016enp}. \\[-.3cm]

The most stringent constraints on cLFV come from the measurement of the branching fractions of the decays $\mu\to e \gamma$, $\mu\to 3 e$ and the conversion rate of muon to electron in the muonic gold.
The current experimental bounds and future sensitivity on these observables are given in \autoref{tab:LFV}. 

It is interesting to mention that, although the current sensitivity of $\mathcal C (\mu, Ti \to e, Ti)$ is $4.3\times10^{-12}$ \cite{SINDRUMII:1993gxf},
which is much weaker than the sensitivity of $\mathcal C (\mu, Au \to e, Au)$, it is expected to reach up to $\mathcal O (10^{-18})$ \cite{prism} in the future. 
Both Mu2e \cite{Mu2e:2014fns} and COMET \cite{COMET:2018auw} are expected to achieve sensitivities of $\mathcal O (10^{-17})$ for $\mu-e$ conversion with Al target.
Concerning cLFV in $\tau$ decay, the current sensitivity of $\mathcal O (10^{-8})$ \cite{Banerjee:2022xuw} for both $\tau\to l \gamma$ and $\tau \to 3 l$,
should be improved by Belle II to $\mathcal O (10^{-9})$ and $\mathcal O (10^{-10})$ respectively, does not substantially constrain the parameter-space of our scotogenic model. \\[-.3cm] 

\begin{figure}[h!]
	\scalebox{1.1}{\hspace{-1cm}
		\includegraphics[scale=0.3]{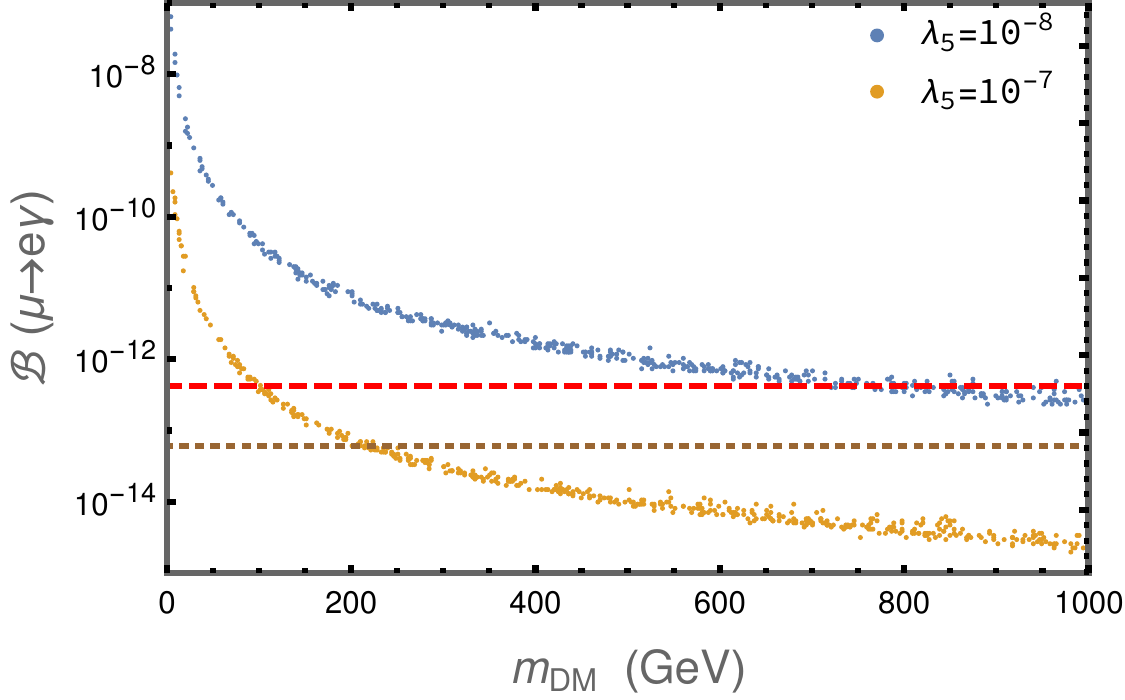}\hfil
		\includegraphics[scale=0.3]{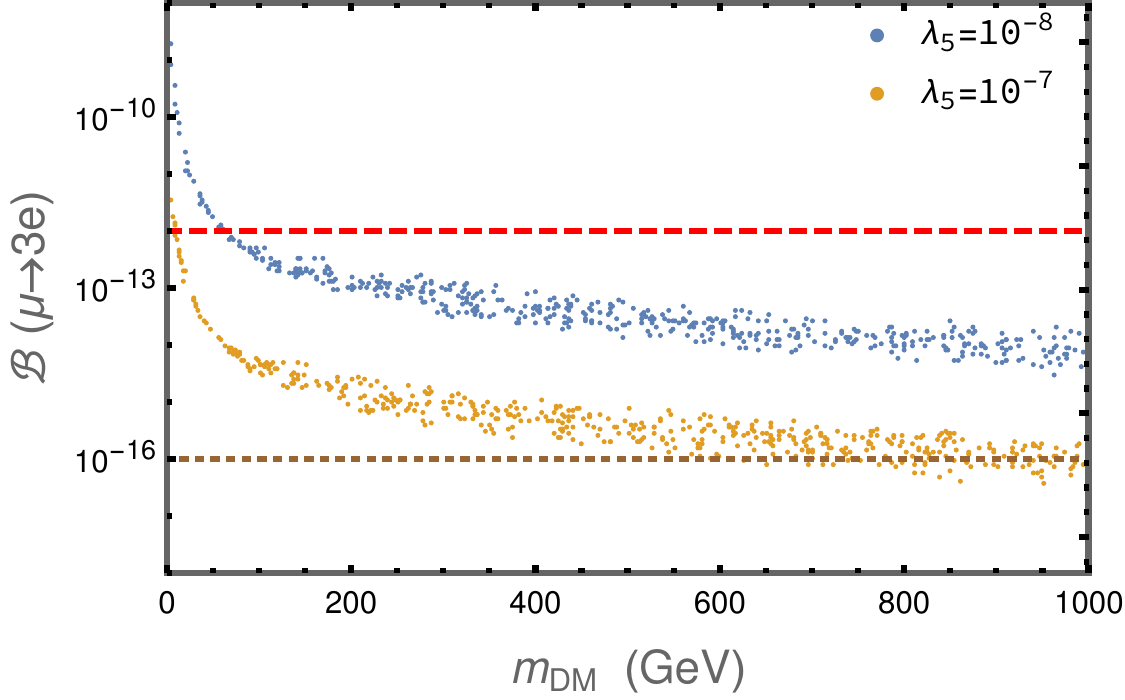}\hfil
		\includegraphics[scale=0.3]{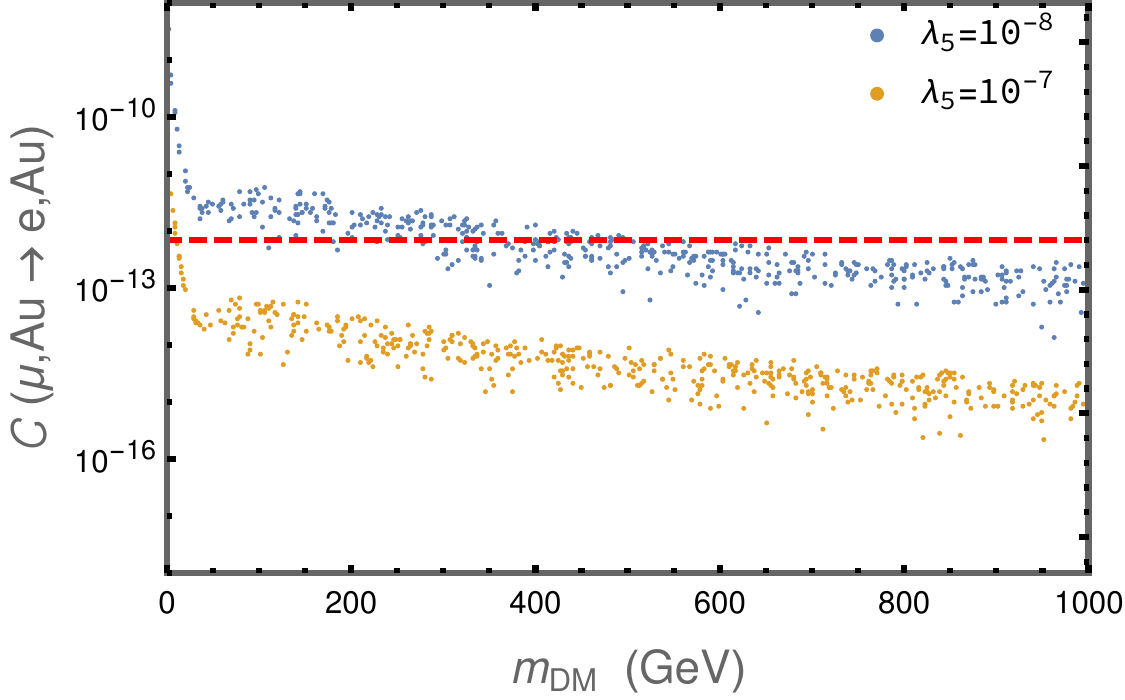}}\\[2mm]
	\vglue 1cm	
	\scalebox{1.1}{\hspace{-1cm}
		\includegraphics[scale=0.3]{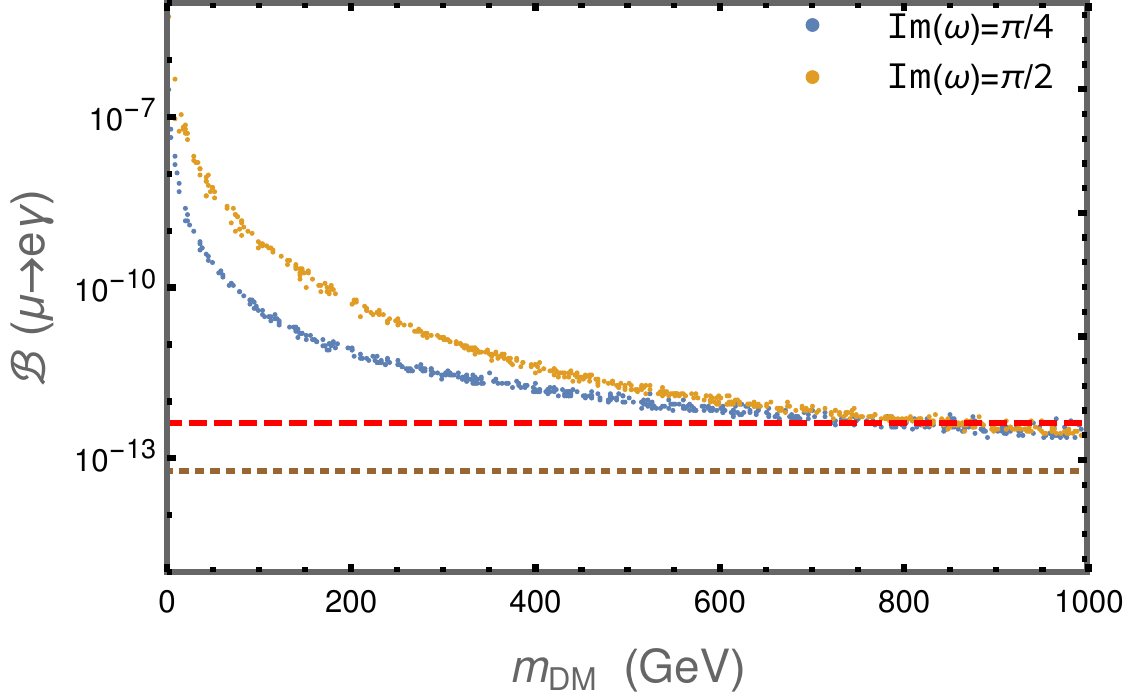}\hfil
		\includegraphics[scale=0.3]{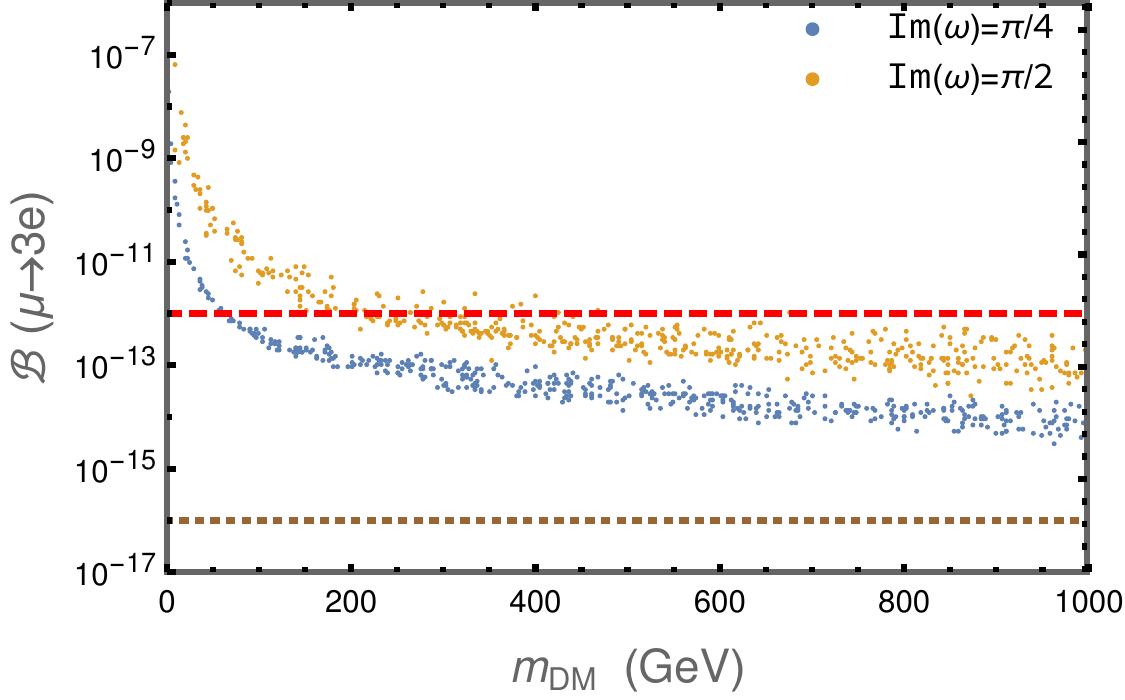}\hfil
		\includegraphics[scale=0.3]{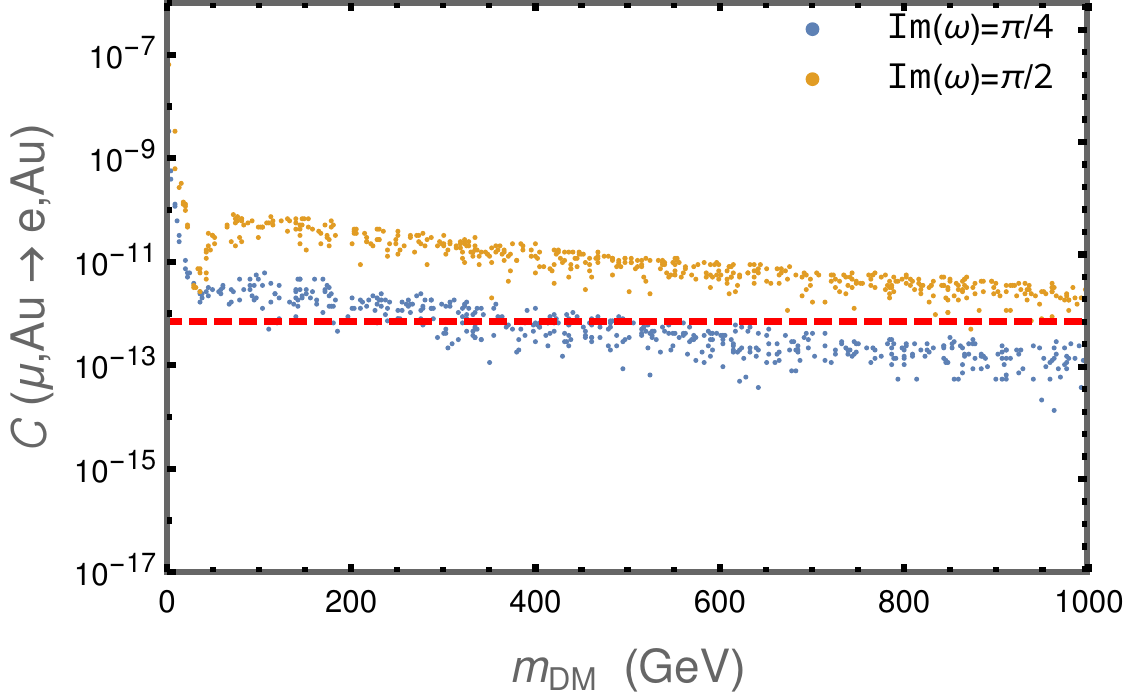}}
	\caption{$\mathcal B (\mu\to e \gamma)$, $\mathcal B (\mu\to 3e)$ and $\mathcal C (\mu, Au \to e, Au)$ versus $\lambda_5$ and $\text{Im}(\omega)$.
		The other parameters are fixed as in BP0 benchmark defined in \autoref{tab:BP0}.
		The blue points in all the plots correspond to BP0, while the orange points are generic.
		The horizontal red dashed lines indicate the current experimental bounds, whereas the brown dotted lines correspond to future sensitivities.}
	\label{fig:LFV}
\end{figure}
In Fig.~\ref{fig:LFV} we display the rates for these cLFV observables. Apart from a mild dependence on the parameters \dsf and \depf which determine $\Sigma^+$ and $\eta^+$ masses running in the loops,
  these processes are mainly controlled by $M_F$, $\lambda_5$ and Im$(\omega)$.
In the top row we present these observables versus the dark matter mass \mdm for two different values of $\lambda_5$: $1.0 \times 10^{-8}$ (blue) and $1.0 \times 10^{-7}$ (yellow). 
The bottom row shows the same for two different values of Im$(\omega)$: $\pi/4$ (blue) and $\pi/2$ (yellow). 
These cLFV observables depend mainly on the Yukawa couplings $Y_F^\alpha$ and $Y_\Sigma^\alpha$, which are governed by $\lambda_5$ and Im$(\omega)$. 
It is interesting to mention that Re$(\omega)$ contributes to the Yukawa couplings as trigonometric `$\sin$' and `$\cos$' functions, with mild variation,
whereas Im$(\omega)$ affects the Yukawa couplings as hyperbolic functions, with a steeper variation. \\[-.3cm] 

In summary, one sees that the attainable values for these cLFV observables can all exceed the current experimental limits for low dark matter mass values.
In fact, $\mu\to e\gamma$ rules out the benchmark point BP0 up to 800 GeV of the dark matter mass.
This fact plays an important role in ruling out the viability of low-mass dark matter in this model.  \\

       Before closing this section, let us also present the current sensitivities of cLFV tau decays, along with their future projections in \autoref{tab:LFVtau}~
      \footnote{For the limits on cLFV semi-leptonic channels, see Ref. \cite{Banerjee:2022xuw}.}.
      For the benchmark values taken above we find that the numerically computed branching ratios for these processes stay well within their current experimental limits.
 Only by taking very small values of $\lambda_5 \le 10^{-10}$, these tau decays become competitive with other cLFV observables, in agreement with the results obtained in Ref.~\cite{Rocha-Moran:2016enp}.
      
 In summary, we conclude that $\rm BR(\mu \to e \gamma), \rm BR(\mu \to 3 e)$ and $\rm Rate( \mu, Au \to e, Au)$ remain as the three main cLFV observables to be explored in our
 fermionic scotogenic dark matter scenario.
However, cLFV tau decays may also provide alternative probes to envisage in the future. \\

\begin{table}[h!]
	\begin{tabular}{| c | c | c |}
		\hline
		\;	LFV Process \; & \; Current Bound \; & \; Future Sensitivity \;\\
		\hline
		$\mathcal B (\tau\to e \gamma)$& $3.3 \times 10^{-8}$ \cite{BaBar:2009hkt,Banerjee:2022xuw} & $9.0 \times 10^{-9}$ \cite{Belle-II:2022cgf,Banerjee:2022xuw}\\
		\hline
		$\mathcal B (\tau\to \mu \gamma)$& $4.2 \times 10^{-8}$ \cite{Belle:2021ysv,Banerjee:2022xuw} & $6.9 \times 10^{-9}$ \cite{Belle-II:2022cgf,Banerjee:2022xuw}\\
		\hline
		$\mathcal B (\tau\to 3e)$& $2.7 \times 10^{-8}$ \cite{Hayasaka:2010np,Banerjee:2022xuw} & $4.7 \times 10^{-10}$ \cite{Belle-II:2022cgf,Banerjee:2022xuw}\\
		\hline
		$\mathcal B (\tau\to 3\mu)$& $2.1 \times 10^{-8}$ \cite{Hayasaka:2010np,Banerjee:2022xuw} & $3.6 \times 10^{-10}$ \cite{Belle-II:2022cgf,Banerjee:2022xuw}\\
		\hline
		$\mathcal B (\tau^-\to e^-\mu^+ \mu^-)$& $2.7 \times 10^{-8}$ \cite{Hayasaka:2010np,Banerjee:2022xuw} & $4.5 \times 10^{-10}$ \cite{Belle-II:2022cgf,Banerjee:2022xuw}\\
		\hline
		$\mathcal B (\tau^-\to e^+\mu^- \mu^-)$& $1.7 \times 10^{-8}$ \cite{Hayasaka:2010np,Banerjee:2022xuw} & $2.6 \times 10^{-10}$ \cite{Belle-II:2022cgf,Banerjee:2022xuw}\\
		\hline
		$\mathcal B (\tau^-\to \mu^- e^+ e^-)$& $1.8 \times 10^{-8}$ \cite{Hayasaka:2010np,Banerjee:2022xuw} & $2.9 \times 10^{-10}$ \cite{Belle-II:2022cgf,Banerjee:2022xuw}\\
		\hline
		$\mathcal B (\tau^-\to \mu^+ e^- e^-)$& $1.5 \times 10^{-8}$ \cite{Hayasaka:2010np,Banerjee:2022xuw} & $2.3 \times 10^{-10}$ \cite{Belle-II:2022cgf,Banerjee:2022xuw}\\
		\hline
	\end{tabular}
	\caption{Experimental bounds on rare leptonic cLFV $\tau$ decay processes. }
	\label{tab:LFVtau}	
\end{table}

\section{Fermionic Dark Matter Phenomenology} 

 In this section, we perform a detailed study of fermionic dark matter in the singlet-triplet scotogenic model, highlighting some major phenomenological aspects. 
 The possibility of scalar scotogenic dark matter was explored previously in Ref.~\cite{Diaz:2016udz,Avila:2019hhv}. 
 Some fermionic dark matter results have also been given in a recent work in Refs.~\cite{Restrepo:2019ilz}.

We briefly discuss two major astrophysical constraints on the dark matter sector: the constraints imposed by the measured relic abundance as well as the direct detection experiments.

\begin{itemize}
\item {\bf Relic Density} \\
 When the dark matter falls out of thermal equilibrium in the early Universe, the remnant dark matter remains frozen out.  
 The relic dark matter density given by the Planck measurement~\cite{Planck:2018vyg} is
\begin{equation}
	\Omega h^2 = 0.120\pm 0.001 .    
\end{equation}

\vspace*{-5mm}

\item {\bf Direct Detection Constraints}\\
  Several experiments have performed direct searches for dark matter by attempting to detect nuclear recoil from scattering of dark matter off given target nuclei. 
In the light mass regime from 0.1 GeV to 4 GeV, the most stringent upper bound on the spin-independent dark matter-nucleon scattering cross-section comes from DarkSide-50 \cite{DarkSide-50:2023fcw}.
  In the window of 4 GeV to 10 GeV, the strongest constraints are by XENON1T $^8$B \cite{XENON:2020gfr} and PandaX-4T $^8$B \cite{PandaX:2022aac}.
  In the range of 10 GeV to 10 TeV, experiments like XENON1T \cite{XENON:2018voc}, XENONnT \cite{XENONCollaboration:2023orw}, PandaX-4T \cite{PandaX-4T:2021bab} impose similar constraints on the
  spin-independent cross-section.  
  Nevertheless, the strongest restriction in this mass range comes from LZ \cite{LUX-ZEPLIN:2022xrq}.  

  Note that the ``coherent scattering of neutrinos'' can also produce nuclear recoil and act as background to the WIMP searches.
  This effect produces the ``neutrino floor'' \cite{Billard:2021uyg,OHare:2021utq}, providing an intrinsic limitation on the discovery of dark matter.
  Thus the direct detection experiments allows a narrow window for the spin independent neucleon-DM cross-section. 
\end{itemize}

In what follows we study the dependences of these two observables on the parameters of the triplet-singlet scotogenic model. 
All the dark matter analyses in this section are performed using BP0 as the benchmark point, using the {\tt micrOMEGAs} tool~\cite{Belanger:2001fz,Belanger:2020gnr}

\subsection{ Cosmological Relic Density } 
\label{sub:relic_a}
\vspace*{-3mm}

\begin{figure}[h!]
	\scalebox{1.1}{
	\hspace{-0.5cm}
	\includegraphics[scale=0.18]{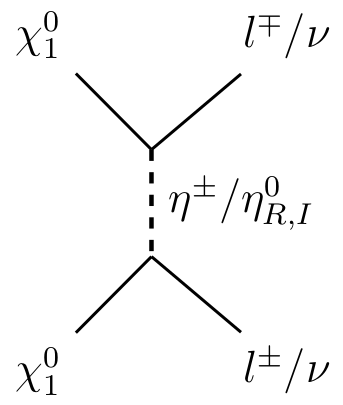}\hfil
	\includegraphics[scale=0.18]{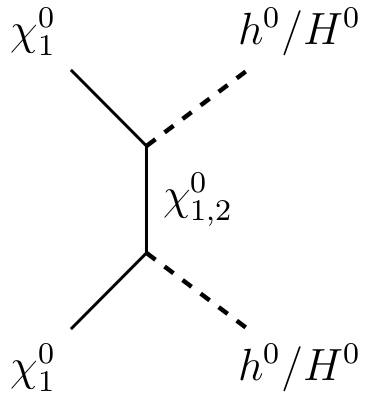}\hfil
	\includegraphics[scale=0.18]{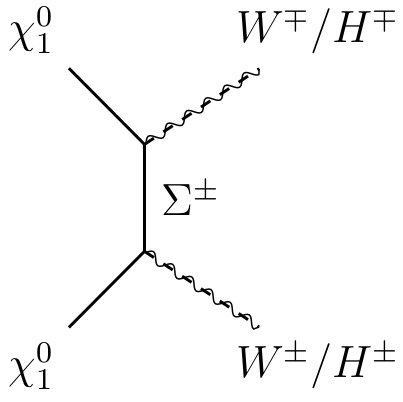}\hfil
	\raisebox{4.0mm}[0pt][0pt]{
	\includegraphics[scale=0.18]{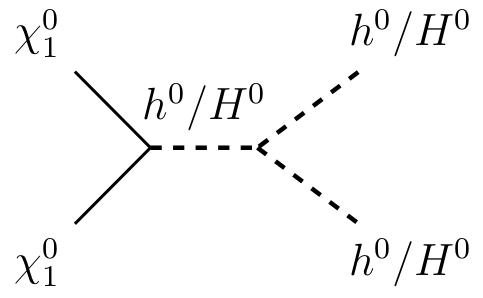}\hfil
	\includegraphics[scale=0.18]{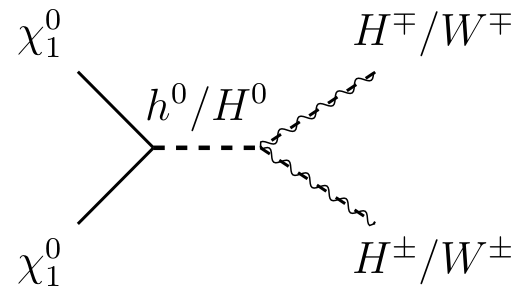}\hfil
	\includegraphics[scale=0.18]{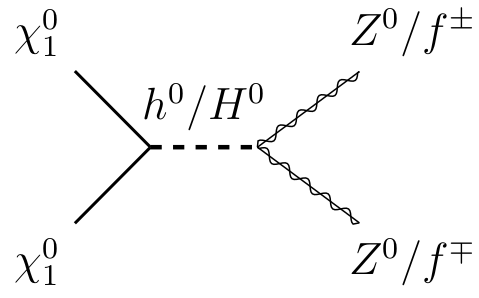}}}
    \caption{The $\eta$, $\chi_{1,2}^0$ and $\Sigma^{\pm}$--mediated fermionic t-channel dark matter annihilation processes. The three s-channels on the right panels
      correspond to resonant annihilation through the neutral $\mathcal Z_2$-even scalars $h^0,H^0$.}
\label{fig:ann-DM}
\end{figure}

The annihilation channels for a fermionic dark matter candidate $\chi^0_1$ in the singlet-triplet scotogenic model are given in~Fig.~\ref{fig:ann-DM}. 
The annihilation processes can be divided into three broad categories, with distinct phenomenological implications for the dark matter sector. 

We now discuss some of their salient features.
\begin{itemize}
\item 
  A pair of fermionic dark matter particles annihilate to visible sector final states, through t-channel (and u-channel) processes mediated by dark scalars $\eta_{R,I}^0, \eta^{\pm}$ and
  fermions $\chi^0_{1,2}, \Sigma^{\pm}$.
  The strength of these annihilation involves Yukawa couplings that determine radiative neutrino mass generation. %, along with other gauge and Yukawa couplings.
  The relevant Feynman diagrams are shown in Fig.~\ref{fig:ann-DM} (first three diagrams).

\item
Another key dark matter annihilation mechanism in this scenario is through scalar resonances.
This includes the Higgs portal as usual, while a triplet-like scalar $H^0$ is also present as a portal of dark matter annihilation.  
The $H^0$ mediated processes happen only due to the presence of $\phi^0-\Omega^0$ mixing.  
The $\chi_1^0 \chi^0_1 \to h^0/H^0 \to SM~SM$ modes have a resonant enhancement at $m_{\chi^0_1} \sim m_{h^0/H^0}/2$. 
The relevant Feynman diagrams are given in Fig.~\ref{fig:ann-DM} (last three diagrams). 

\item 
  Another dark matter annihilation channel takes place when there is a mass degenerate dark sector.
 This can happen due to the presence of dark particles only marginally heavier than the lighest one. 
In this case the other particles in the nearly degenerate dark sector take part in the annihilation, hence termed as dark matter co-annihilation. 
Dark matter co-annihilation was first discussed in Ref.~\cite{Griest:1990kh}. 
The thermal--averaged dark matter co-annihilation cross section is parametrized as $$ \vev{ \sigma_c v } \propto \sum_{i,j} \sigma_{ij} e^{-\frac{m_i - m_{\rm DM}}{T}}  e^{-\frac{m_j - m_{\rm DM}}{T}}. $$ 
In our case the relevant Feynman diagrams are given in Appendix~\ref{sec:co}, see~Fig.~\ref{fig:ffco} and~Fig.~\ref{fig:fsco}. 
\end{itemize}

 %%%%%%%%%%%%%%%%%%%%%%%%%%%%%%%%%%%%%%%%%%%%%%%%%%%%%%%%%%%%%%%%%%%%%%%%%%%%%%%%%%%%%%%%%
 
 \begin{figure}[h!]
 	\includegraphics[scale=0.4]{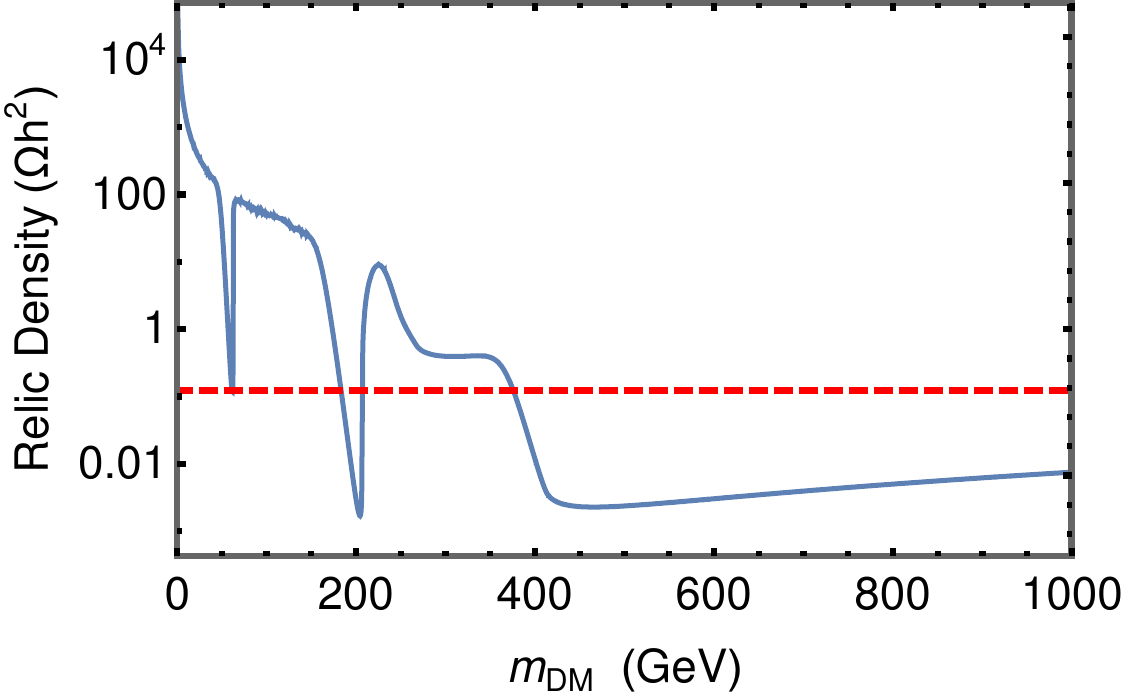}
 	\caption{Relic density versus dark matter mass for the benchmark point BP0. The red dashed line indicates the observed Planck value.}
 	\label{fig:relic0}
 \end{figure}

 In Fig.~\ref{fig:relic0} we depict the variation of the dark matter relic density with its mass for the benchmark point BP0. 
 Since \dsf and \depf are large in BP0 (see \autoref{tab:BP0}), co-annihilation is negligible, so the relic density is fully determined by the diagrams shown in Fig.~\ref{fig:ann-DM}.
 As the dark matter mass increases, different annihilation channels open up, and the relic density decreases gradually, as can be seen in Fig.~\ref{fig:relic0}. 
 One can notice two sharp resonant dips at 62.5 GeV and 200 GeV (i.e. $m_{h^0}^{}/2$ and $m_{H^0}^{} /2$).
 These correspond to s-channel dark matter annihilation through the neutral scalars $h^0$ and $H^0$ respectively, i.e. the processes $\chi_1^0 \chi_1^0 \to h^0/H^0 \to SM~SM$.
 Moreover, at $m_{\rm DM}^{}=(m_{W^\pm}^{}+m_{H^\pm}^{})/2\approx242$ GeV and $m_{\rm DM}^{}=(m_{h^0}^{}+m_{H^0}^{})/2\approx263$ GeVs respectively, two new annihilation channels
 $\chi_1^0 \chi_1^0 \to W^\pm H^\pm$ and $\chi_1^0 \chi_1^0 \to h^0 H^0$ open up, reducing the relic density. 
 This corresponds to the fall-off in the region around $\sim 250$~GeV.
 Another sharp drop is seen at 400 GeV, illustrating the opening of the phase space for  $\chi_1^0 \chi_1^0 \to H^0 H^0$ and $\chi_1^0 \chi_1^0 \to H^+ H^-$ respectively. 
 Beyond 400 GeV, the relic abundance of $\chi_1^0$ is mainly controlled by the annihilation channels to $H^0 H^0$ and $H^+ H^-$.

\begin{figure}[h!]
 \scalebox{1.1}{\hspace{-0.5cm}\includegraphics[scale=0.3]{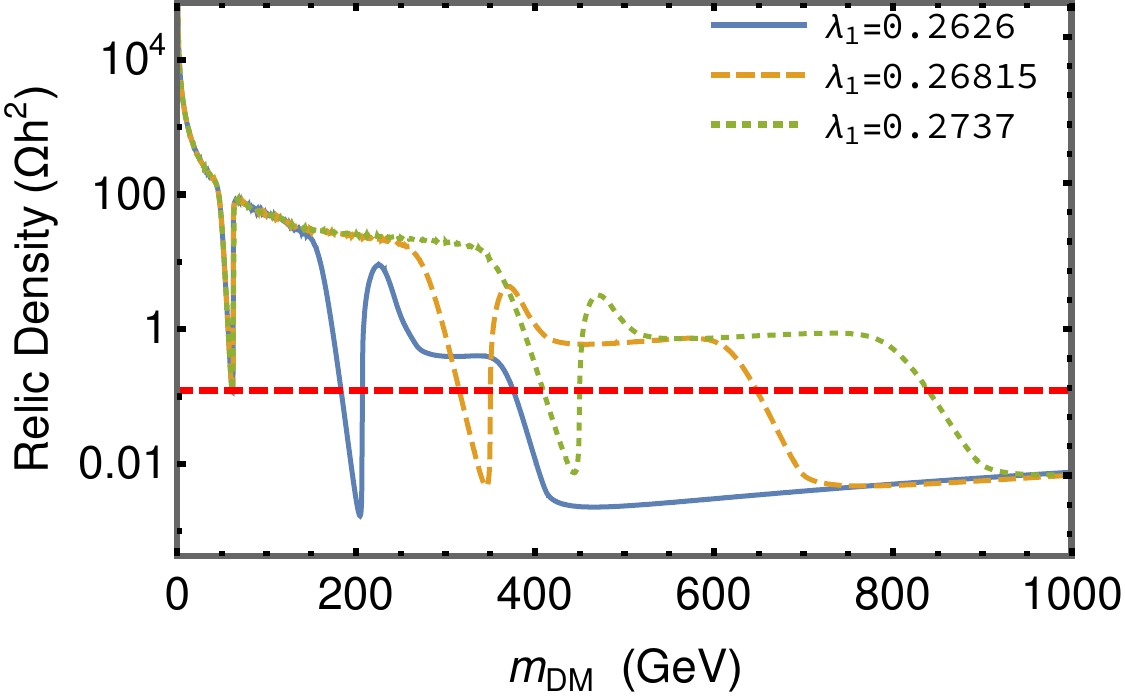}\hfil	
 	\includegraphics[scale=0.3]{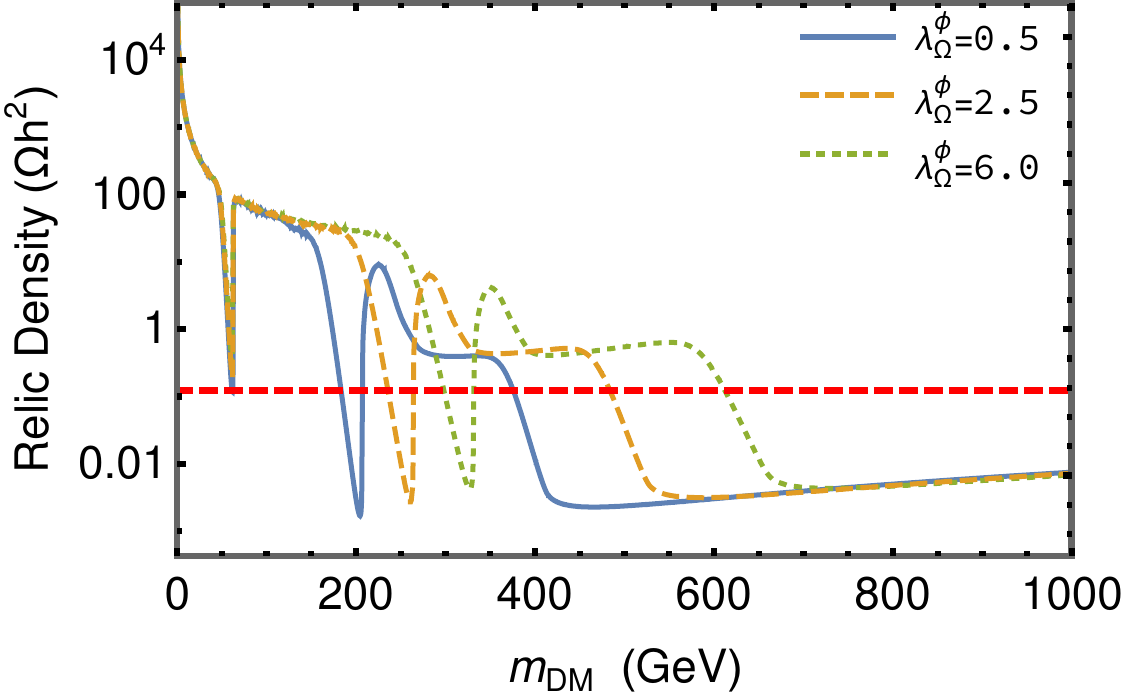}\hfil
 	\includegraphics[scale=0.3]{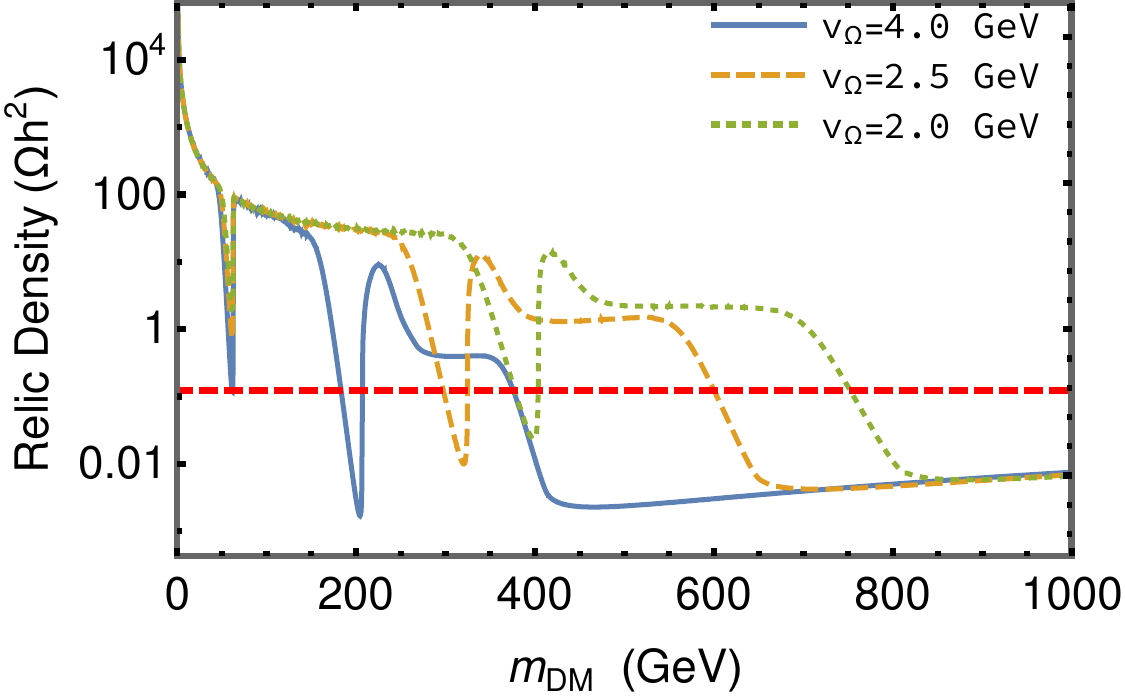}}
      \caption{Relic density versus dark matter mass for different $\lambda_1$, $\lambda^{\phi}_\Omega$ and $v_\Omega$ with the remaining parameters fixed as in BP0, \autoref{tab:BP0}.
        These parameters control the mass of the second neutral scalar $H^0$. The blue curve represents BP0. The red dashed line indicates the observed Planck value.}
 \label{fig:relic1}
\end{figure}

 We now turn to the dependencies of the dark matter relic density on the parameters of our model.
 As we saw in the above paragraph, the relic density is very sensitive to the mass of the heavy Higgs particles, $m_{H^0} \approx m_{H^\pm}$.  
 In our parametrization the value of $m_{H^0}$ mainly depends on three \footnote{Though  Eq. \eqref{Eq:ScalarMM} suggests additional dependence of $m_{H^0}$ on $\mu_{\Omega}^{\phi}$ and $\lambda_{\Omega}^{\Omega}$, we note that $\mu_{\Omega}^{\phi}$
   is not an independent parameter, while $\lambda_{\Omega}^{\Omega}$ comes with $v_\Omega$, making its contribution negligible.} parameters: $\lambda_1, \lambda_{\Omega}^{\phi}, v_{\Omega}$.
In the three panels of Fig.~\ref{fig:relic1} we present the individual impact of these three parameters on the relic density. 
 We choose three different values for each of these parameters, keeping the undisplayed parameters fixed as BP0. 
 One sees that, except the first dip at $m_{h^0}/2$, all others present in Fig.~\ref{fig:relic1} shift towards the right as $\lambda_1$ or $\lambda_\Omega^{\phi}$ increases,
 or as $v_\Omega$ decreases, corresponding to a larger value of $m_{H^0}$. 
 The most sensitive parameter dependence however, appears to be on $\lambda_1$. 
 With increasing $m_{H^0}$, the relic density dips at $m_{\rm DM} \sim m_{H^0}/2$ move upwards,
 as the annihilation at resonance gradually decreases with increasing ${H^0}$ mass and its Breit-Wigner width.

\begin{figure}[h!]
  \scalebox{1.1}{\hspace{-0.5cm}\includegraphics[scale=0.3]{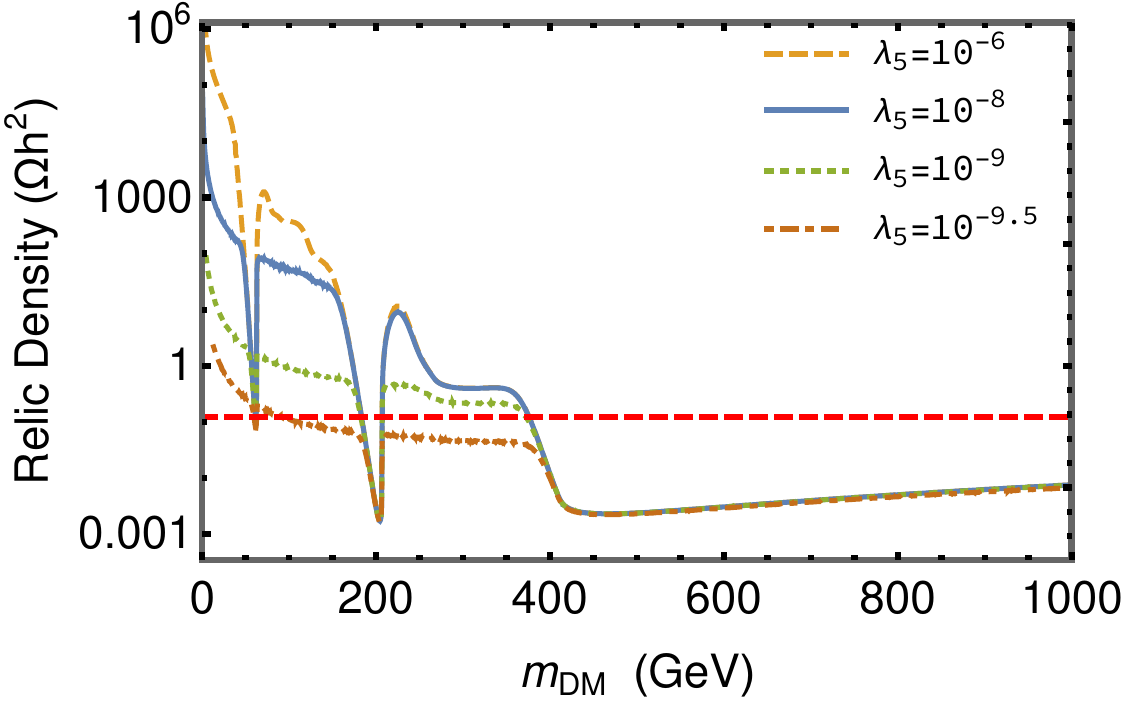}\hfil	
 	\includegraphics[scale=0.3]{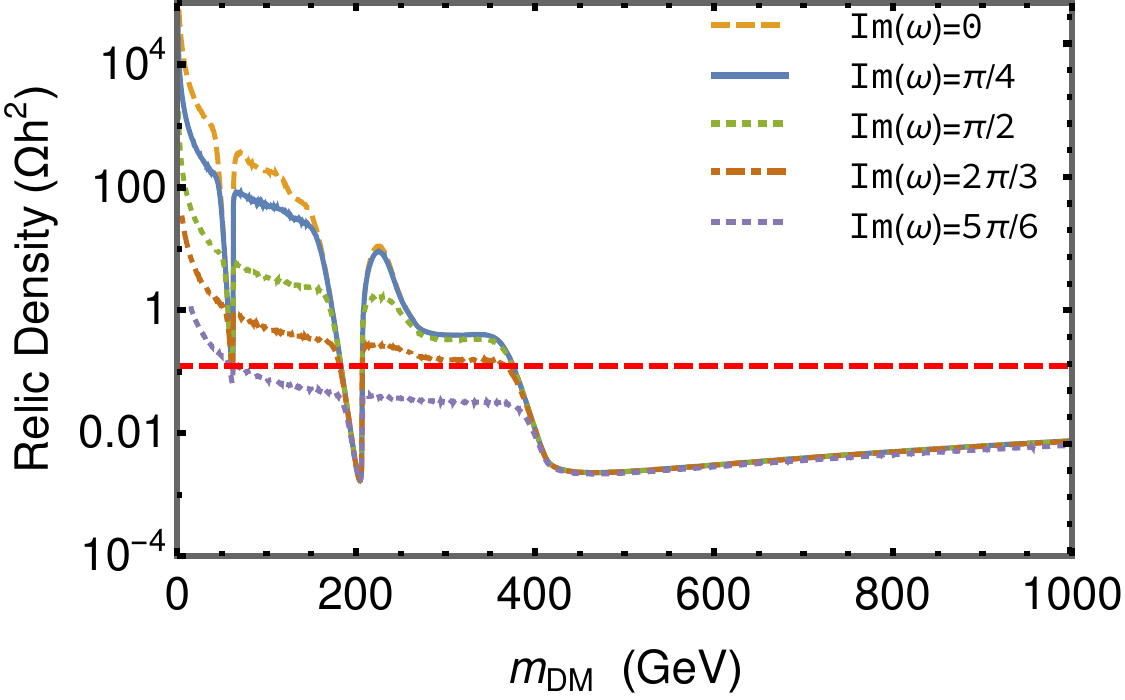}\hfill\includegraphics[scale=0.3]{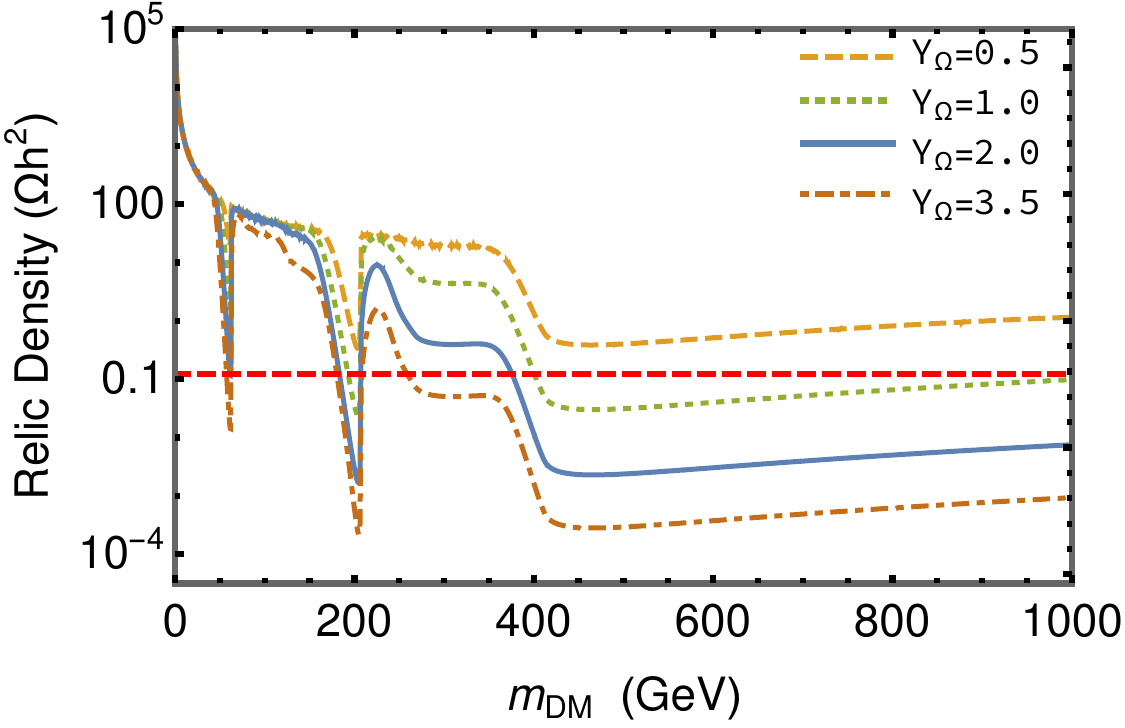}}
      \caption{Relic density versus dark matter mass for different $\lambda_5$, $\text{Im}(\omega)$ and $Y_\Omega$ fixing the other parameters as in BP0, \autoref{tab:BP0}.
        The blue curve represents BP0. (The parameter $\text{Re}(\omega)$ has very minor effects on the relic density). The red dashed line indicates the observed Planck value.}
 	\label{fig:relic2}
 \end{figure}

The other important parameters governing the relic density are the Yukawa couplings $Y_F$, $Y_\Sigma$ and $Y_\Omega$. 
While the first t-channel diagram in Fig.~\ref{fig:ann-DM} depends on $Y_F$ and $Y_\Sigma$, the last three diagrams depend on the coupling $Y_\Omega$. 
In order to satisfy the neutrino oscillation data, the Yukawa couplings $Y_F$ and $Y_\Sigma$ are mainly determined by $\lambda_5$ and Im$(\omega)$.
Therefore, in Fig.~\ref{fig:relic2} we present the dependece of the relic density density on $\lambda_5$, Im$(\omega)$ and $Y_\Omega$. 
In the first two panels of Fig.~\ref{fig:relic2} one sees that by decreasing $\lambda_5$ or increasing Im$(\omega)$ one enhances $Y_F$ and $Y_\Sigma$, which in turn reduce the relic density.
Since these couplings do not affect the $H^0 H^0$ and $H^+ H^-$ channels that determine the relic density for $m_{\rm DM}> m_{H^0}$, these Yukawas do not affect the relic density in that region. 
In contrast, one sees from the third panel in Fig.~\ref{fig:relic2} that the relic density decreases sharply with increasing $Y_\Omega$ for higher DM mass. 
This happens since the parameter $Y_\Omega$ determines the couplings for $\chi_i^0 \chi_j^0 H^0$, $\chi_i^0\chi_j^0 h^0$ and $\chi_i^0\Sigma^+H^-$ interactions.
Note however, that the positions of the dips do not shift in Fig.~\ref{fig:relic2}, as the mass $m_{H^0}$ does not depend on these three parameters. 

\begin{figure}[h!]
 \scalebox{1.1}{\includegraphics[scale=0.35]{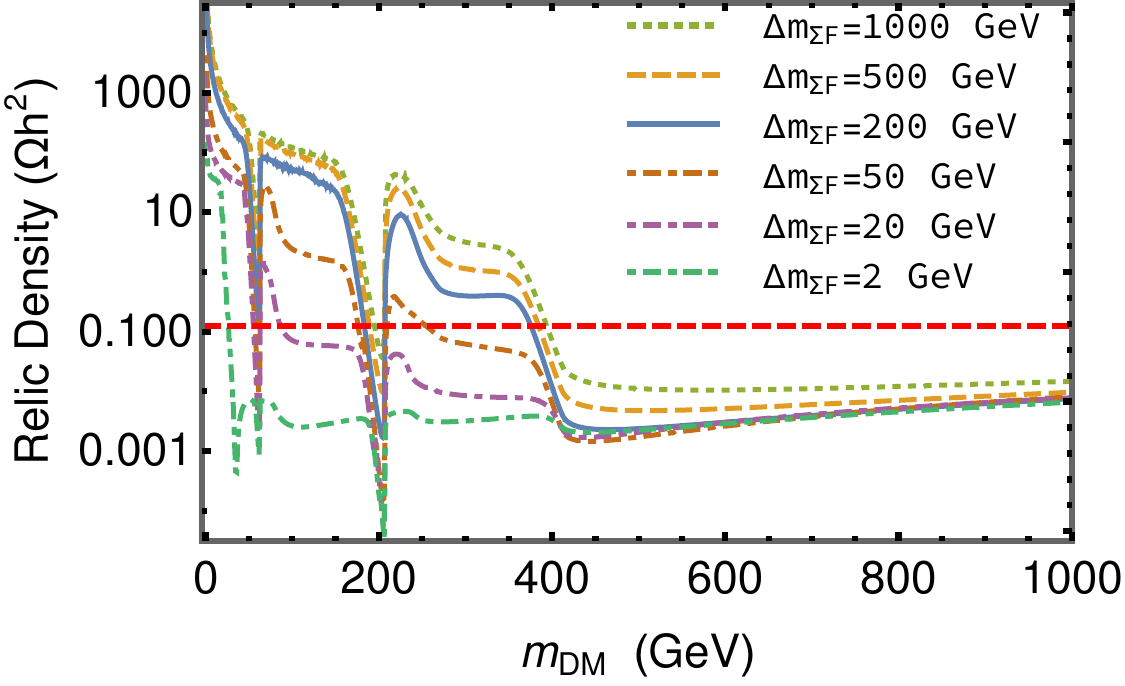}\hspace{5mm}	
 	\includegraphics[scale=0.345]{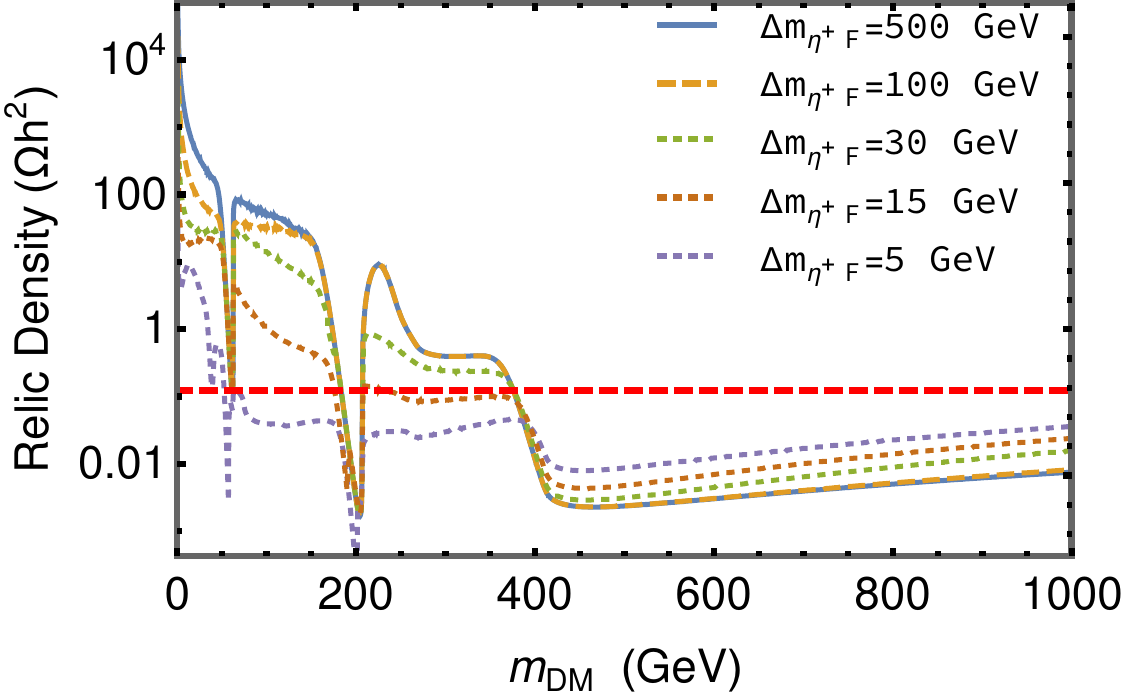}}
      \caption{Relic density versus dark matter mass for different mass-difference parameters $\Delta m_{\Sigma F}$, $\Delta m_{\eta^+ F}$ 
          fixing the other parameters as in BP0, \autoref{tab:BP0}.
          The blue curve represents BP0.
          Fermion-fermion as well as scalar-fermion co-annihilation effects are visible for smaller values of these parameters. The red dashed line indicates the observed Planck value.}
 	\label{fig:relic3}
 \end{figure}

 The remaining parameters that affect the relic density are the mass-difference parameters \dsf and \depf, as illustrated in Fig.~\ref{fig:relic3}.
 % %
 When \dsf and \depf are very small, i.e. the dark sector becomes almost mass degenerate, 
 the dark matter annihilation is assisted by the other dark particles, significantly enhancing the annihilation cross section.
 This reduces the relic density significantly in the region $m_{\rm DM}< m_{H^0}$, due to the activation of new fermion-fermion (Fig.~\ref{fig:ffco}) and
 fermion-scalar (Fig.~\ref{fig:fsco}) co-annihilation channels.
 However, the positions of the dips does not change. 
 
 %%%%%%%%%%%%%%%%%%%%%%%%%%%%%%%%%%%%%%%%%%%%%%%%%%%%%%%%%%%%%%%%%%%%%%%%%%%%%%%%%%%%
 
\vspace*{-7mm}
\subsection{Direct Detection Prospects}
\begin{figure}[h!]
	\centering
	\includegraphics[scale=0.2]{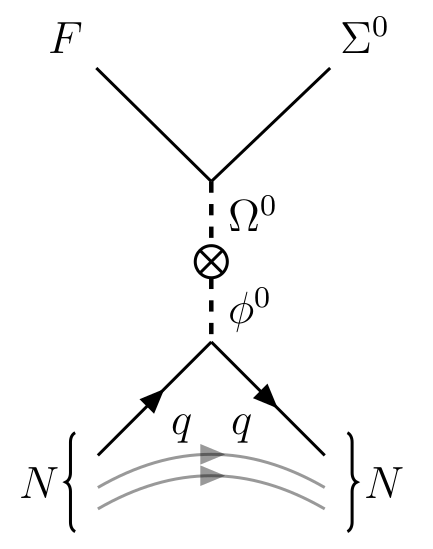}
	\caption{Direct detection process diagram for singlet-triplet mixed fermionic dark matter $\chi_1^0$.
          The Feynman diagram is presented in terms of before mixing fermionic singlet F and triplet $\Sigma$.}
	\label{fig:DD}
\end{figure}

\vspace*{-3mm}
In the triplet-singlet scotogenic model the spin-independent nucleon-dark matter scattering occurs through a t-channel diagram mediated by the neutral non-dark scalars $h^0$ and $H^0$. 
As seen in Fig.~\ref{fig:DD}, in order for this nucleon recoil process to take place, we need both the mixing in the $\mathcal Z_2$-even scalar sector ($\phi^0$ and $\Omega^0$)
as well as in the $\mathcal Z_2$-odd sector ($F$ and $\Sigma^0$). 

 In the limit of negligible Mandelstam variable $t$, this cross-section can easily be estimated as \cite{Restrepo:2019ilz,Choubey:2017yyn}:
 \begin{align}
\label{eq:DD_cr}
\sigma^{\text{SI}}_{\rm DM-N}\approx \frac{\mu_{\text{red}}^2}{\pi}\Big[\frac{Y_\Omega f_N m_N^{}}{2v} \sin2\theta\, \sin2\beta\,\Big(\frac{1}{m_{h^0}^2}-\frac{1}{m_{H^0}^2}\Big)\Big]^2,
 \end{align}
 where $m_N^{}$ is the mass of nucleon, $f_N (\approx 0.3)$ is the nucleon form factor and $\mu_{\text{red}}$ is the reduced mass of nucleon-dark matter system,
 $\displaystyle\mu_{\text{red}}=\Big(\frac{m_{\chi_1^0}\, m_N^{}}{m_{\chi_1^0}+ m_N^{}}\Big)$.  
 Here the parameter $\theta$ is the fermionic mixing angle defined in  Eq. \eqref{eq:tht_mch}, whereas $\beta$ is the scalar mixing angle, expressed as
 \begin{equation}
 	\label{eq:beta}
 	\tan 2\beta=
 	\frac{4v_\Omega v_\phi(\mu_\Omega^\phi-\sqrt 2\lambda_\phi^\Omega v_\Omega)}{4\sqrt 2\lambda_\Omega^\Omega v_\Omega^3-2\sqrt 2\lambda_1 v_\Omega v_\phi^2+ \mu_\Omega^\phi v_\phi^2}.
 \end{equation}

 The spin-independent dark matter nucleon scattering cross section, given by  Eq. \eqref{eq:DD_cr}, depends on the parameters
 $$ \lambda_1, \lambda_{\Omega}^\phi, v_\Omega, \Delta m_{\Sigma F}, Y_\Omega.$$  
 Clearly, the nuclear recoil cross section goes as $\sigma^{\text{SI}}_{\rm DM-N} \propto Y_\Omega^2$.
Moreover, the first three parameters govern the heavy Higgs ($H^0$) mass as well as the scalar mixing angle $\beta$, while the others control the mixing angle $\theta$. 
It is worth mentioning that the direct detection cross section is almost independent of $m_{\rm DM}$, except in the region of light dark matter.
This is evident from  Eq. \eqref{eq:DD_cr}, where $\sigma^{\text{SI}}_{\rm DM-N}$ depends on $m_{\chi_1^0}$ only through $\mu_{\rm red}$, which reduces to $m_N^{}$ for large $m_{\chi_1^0}$ values.

\vspace{-0.6cm}

\section{Dark Matter Phenomenology: results} 
\vspace{-0.2cm}

We now perform a thorough parameter scan in this model, aiming to extract its most salient features concerning fermionic dark matter.
The lightest of the dark triplet-singlet scotogenic fermion mediator will be our fermion dark matter candidate $\chi_1^0$, mainly a singlet. 
The triplet-like dark fermion $\chi^0_2$ can potentially assist the co-annihilation processes. 
Since $v_{\Omega}$ in this model plays a crucial role in the mixing of both scalars and fermions, we broadly divide our analysis in two cases characterized by different $v_{\Omega}$ values.
We dub these as Scenario-I and Scenario-II, respectively, with the parameters specified as in \autoref{tab:BP12}. 

\renewcommand{\arraystretch}{1.5}
\begin{table}[h!]
	\scalebox{0.9}{
		\begin{tabular}
			{||c||c||c|c|c|c||c|c||c|c||c|c|c|c|c|c|c||}
			\hline
			\multirow{2}{*}{Cases}&$v_\Omega$&$M_F$& $\Delta m_{\Sigma F}$& $\Delta m_{\eta^+ F}$& $\Delta m^2_{\eta_I^0\eta^+}$&$\mu_\Omega^\eta$& \multirow{2}{*}{$Y_\Omega$} &\multirow{2}{*}{Re$(\omega)$}&\multirow{2}{*}{Im$(\omega)$}&\multirow{2}{*}{$\lambda_1$}&\multirow{2}{*}{$\lambda_2$}&\multirow{2}{*}{$\lambda_3$}&\multirow{2}{*}{$\lambda_5$}&\multirow{2}{*}{$\lambda_\Omega^\phi$}&\multirow{2}{*}{$\lambda_\Omega^\Omega$}&\multirow{2}{*}{$\lambda_\Omega^\eta$}\\
			&(GeV)&(GeV)&(GeV)&(GeV)&(GeV$^2$)&(GeV)&&&&&&&&&&\\
			\hline\hline
			\multicolumn{17}{||l||}{\textbf{Scenario-I}}\\
			\hline
			BP$_1$&\multirow{3}{*}{\red\framebox{{4.0}}}&\multirow{3}{*}{[3, 10000]}&[100, 500]&[100, 500]&1000&\multirow{3}{*}{400}&\multirow{3}{*}{[0.1, 3.5]}&\multirow{3}{*}{$[-\pi, \pi]$}&\multirow{3}{*}{$[-2\pi, 2\pi]$}& \multirow{3}{*}{0.2626}&\multirow{3}{*}{0.5}&\multirow{3}{*}{0.5}&\multirow{3}{*}{$[10^{-9},0.5]$}&\multirow{3}{*}{0.5}&\multirow{3}{*}{0.5}&\multirow{3}{*}{0.5}\\
			BP$_1^{FF}$&&&[1, 50]&[100, 500]&1000&&&&& &&&&&&\\
			BP$_1^{FS}$&&&[100, 500]&[1, 30]&[1, 1000]&&&&& &&&&&&\\
			\hline\hline
			\multicolumn{17}{||l||}{\textbf{Scenario-II}}\\
			\hline
			BP$_2$&\multirow{3}{*}{\red\framebox{{1.5}}}&\multirow{3}{*}{[3, 10000]}&[100, 500]&[100, 500]&1000&\multirow{3}{*}{400}&\multirow{3}{*}{[0.1, 3.5]}&\multirow{3}{*}{$[-\pi, \pi]$}&\multirow{3}{*}{$[-2\pi, 2\pi]$}& \multirow{3}{*}{0.2626}&\multirow{3}{*}{0.5}&\multirow{3}{*}{0.5}&\multirow{3}{*}{$[10^{-9},0.5]$}&\multirow{3}{*}{0.5}&\multirow{3}{*}{0.5}&\multirow{3}{*}{0.5}\\
			BP$_2^{FF}$&&&[1, 50]&[100, 500]&1000&&&&& &&&&&&\\
			BP$_2^{FS}$&&&[100, 500]&[1, 30]&[1, 1000]&&&&& &&&&&&\\
			\hline
	\end{tabular}}
      \caption{Specifications for Scenario-I ($v_{\Omega}=4.0$~GeV) and Scenario-II ($v_{\Omega}=1.5$~GeV).}
	\label{tab:BP12}
      \end{table}
      
While scanning for each scenario, three different situations are identified, depending on the assumptions made on the dark particle mass-differences, as follows: 
(i) no co-annihilation (both $\Delta m_{\Sigma F}$ and $\Delta m_{\eta^+ F}$ are large, denoted with no superscript),
(ii) fermion-fermion co-annihilation ($\Delta m_{\Sigma F}$ is small but  $\Delta m_{\eta^+ F}$ is large, denoted with superscript `FF')
and (iii) fermion scalar co-annihilation ($\Delta m_{\Sigma F}$ is large but $\Delta m_{\eta^+ F}$ is small, indicated by superscript `FS'). 
The relevant parameters for each of these cases are specified in \autoref{tab:BP12}, and their range of variation is shown as numbers inside square brackets.  
The values of $\lambda_1$, $\lambda_{\Omega}^\phi$ and $v_\Omega$ are kept fixed, in order to fix the mass of the heavy Higgs $H^0$ throughout the scan. 
For definiteness we fix the values of the remaining parameters e.g. $\mu^\eta_\Omega$, $\lambda_2$, $\lambda_3$, $\lambda^\Omega_\Omega$ and $\lambda^\eta_\Omega$ 
as they hardly affect the results. \\[-.3cm]

First we implement the restrictions from Sec.~\ref{sec:constr} including those from neutrino oscillations, encoded in the associated neutrino observables, 
along with the relevant theoretical constraints. 
It is interesting to note that once we adopt the parameterization for the Yukawa couplings $Y_F$ and $Y_\Sigma$ given in  Eq. \eqref{eq:casas},
all the parameter-points generated lie automatically in the region close to the solution of the neutrino oscillation restrictions. 
We then implement the constraints arising from collider physics, cLFV and the measured relic density, successively.  
Finally there are points with under-abundant relic density obeying all of the bounds mentioned.  
Amongst these points, we collect those providing relic density of 1\% to 100\% of the required Planck value in order to study their direct detection prospects.
We present the corresponding results for all the cases mentioned in \autoref{tab:BP12}, which capture the essential features, as follows.\\[-10mm]

\subsection{ Relic Density }
\label{sec:relic-density-}
\vspace*{-3mm}

\begin{figure}[h!]
	\scalebox{1.2}{\hspace{-1cm}
		\includegraphics[scale=0.13]{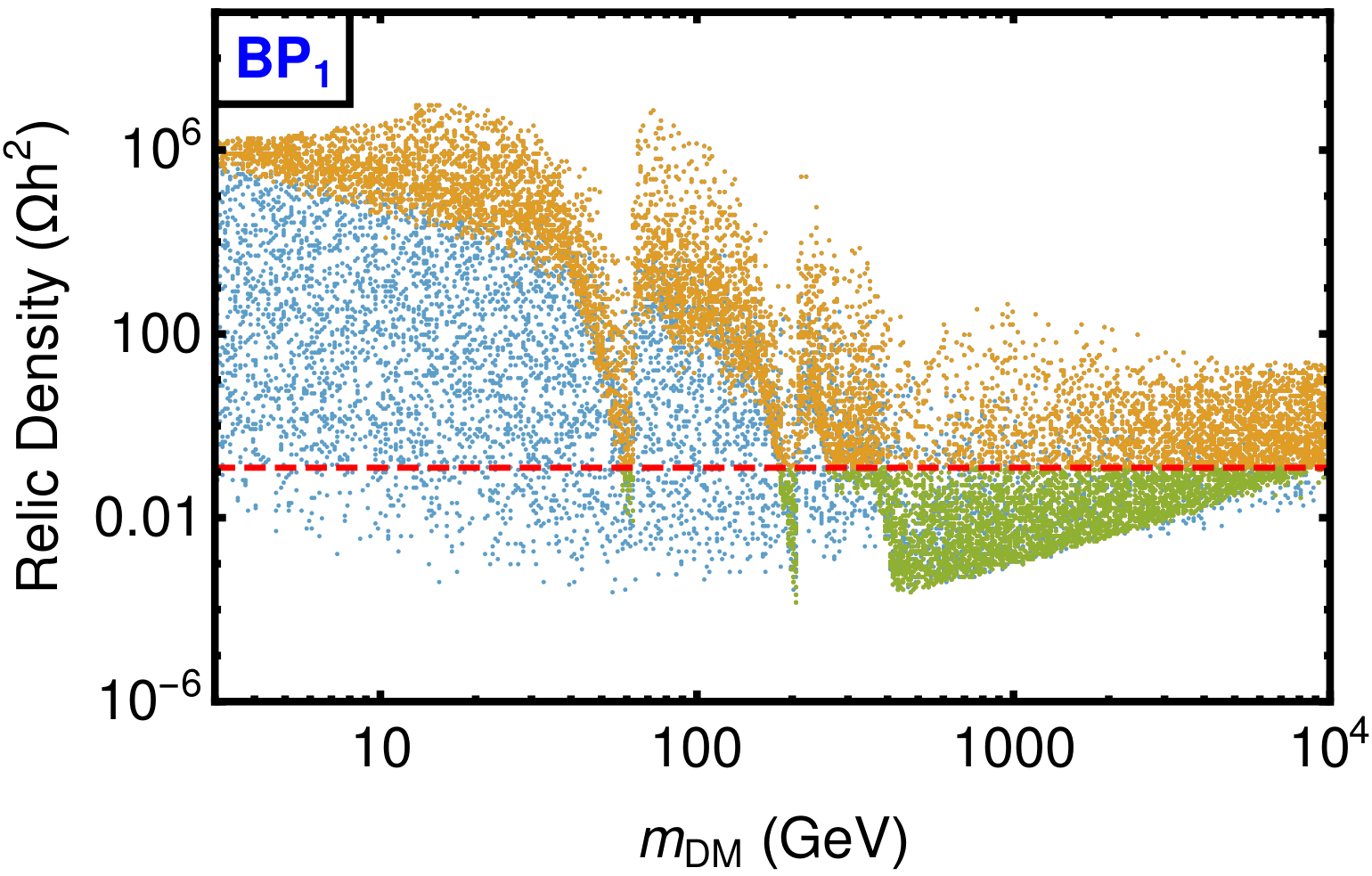}
		\hfill
		\includegraphics[scale=0.13]{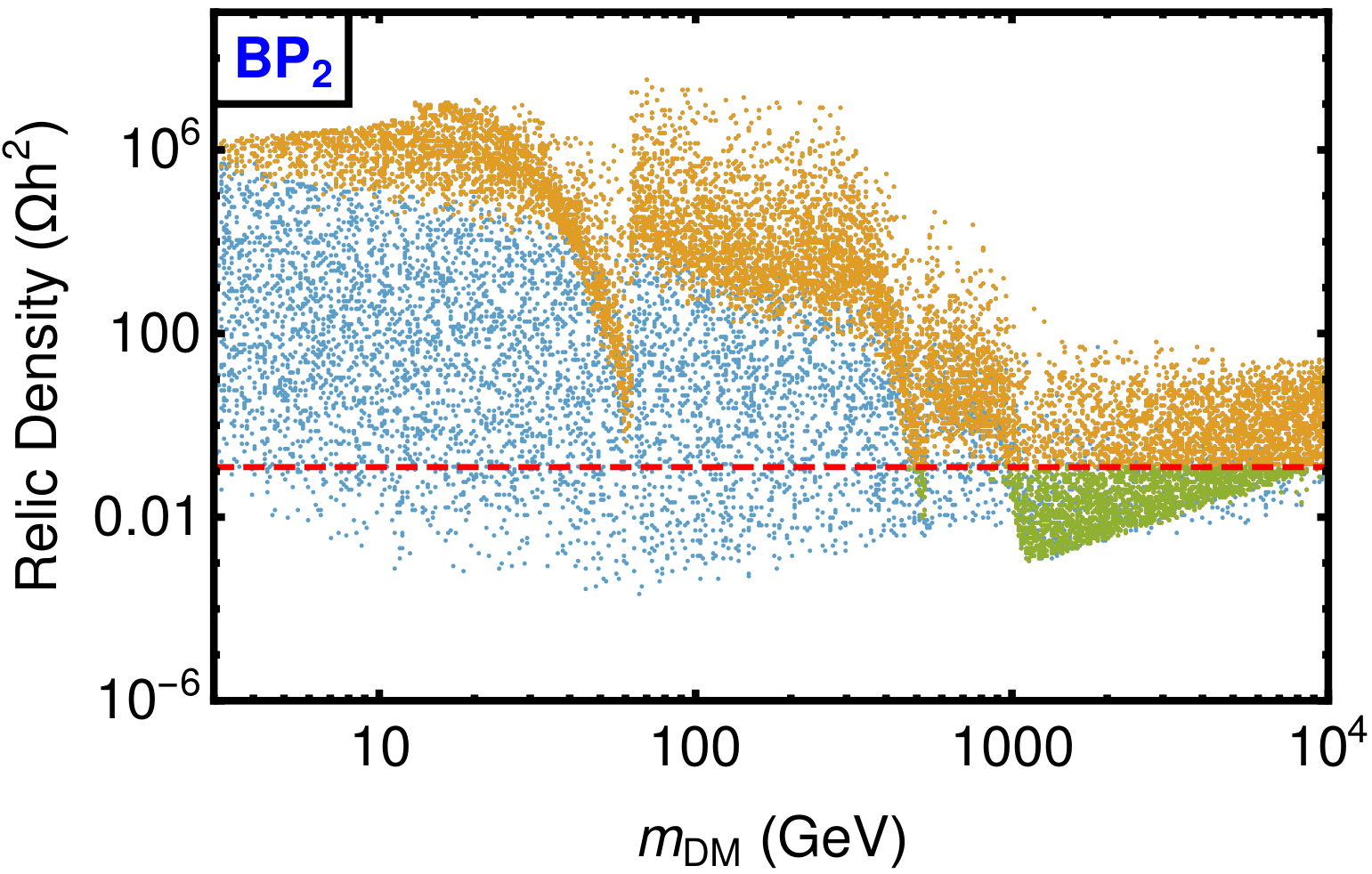}}
	\includegraphics[scale=0.4]{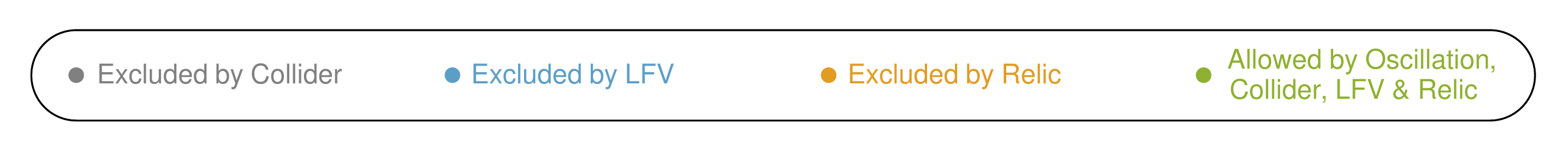}
	\caption{Relic density versus dark matter mass for the no co-annihilation case in Scenario-I (BP$_1$) and Scenario-II (BP$_2$). The red dashed line indicates the Planck value.}
	\label{fig:relicnoco}
\end{figure}

\begin{figure}[h!]
	\scalebox{1.2}{\hspace{-1cm}
		\includegraphics[scale=0.13]{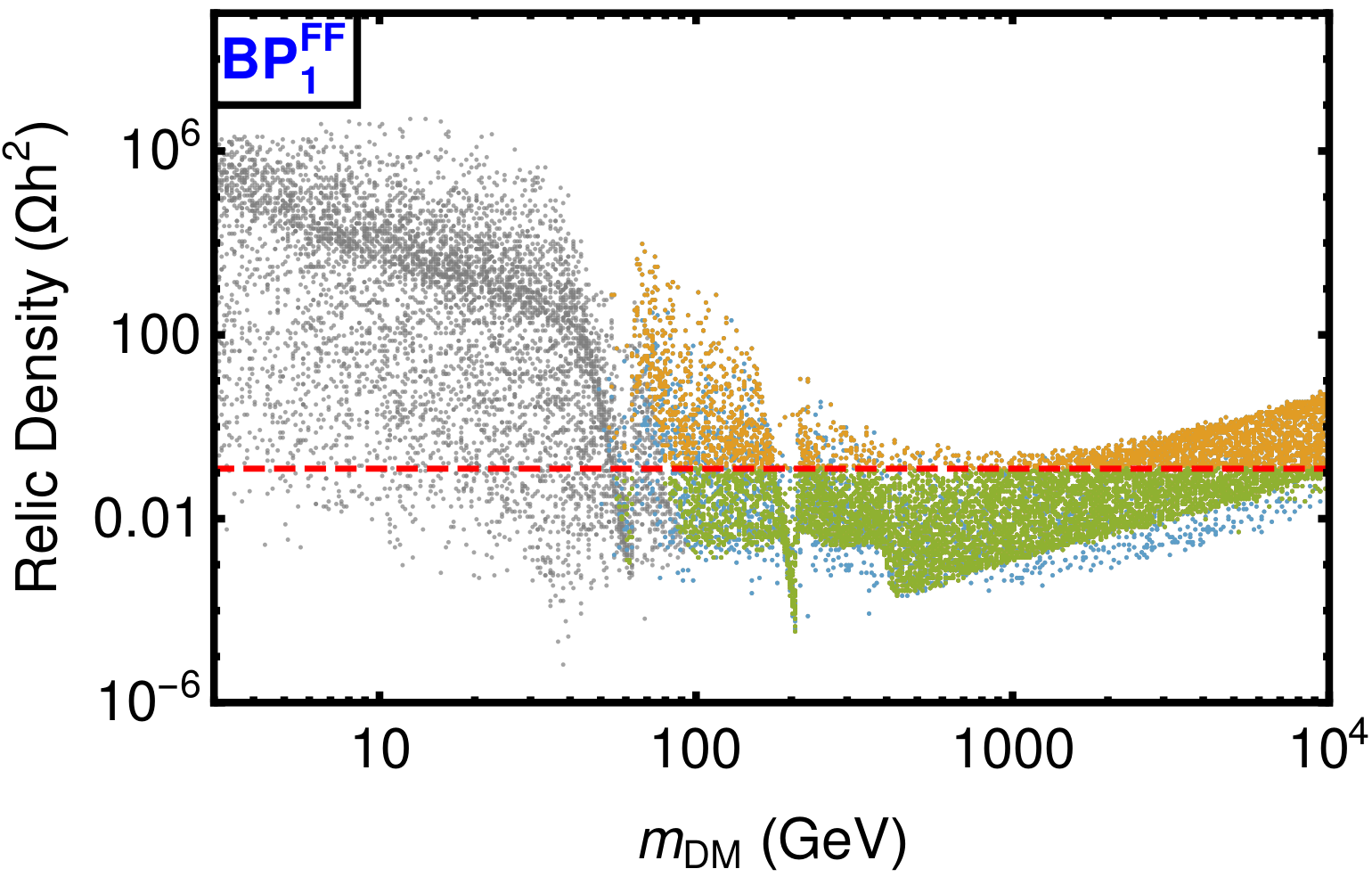}
		\hfill
		\includegraphics[scale=0.13]{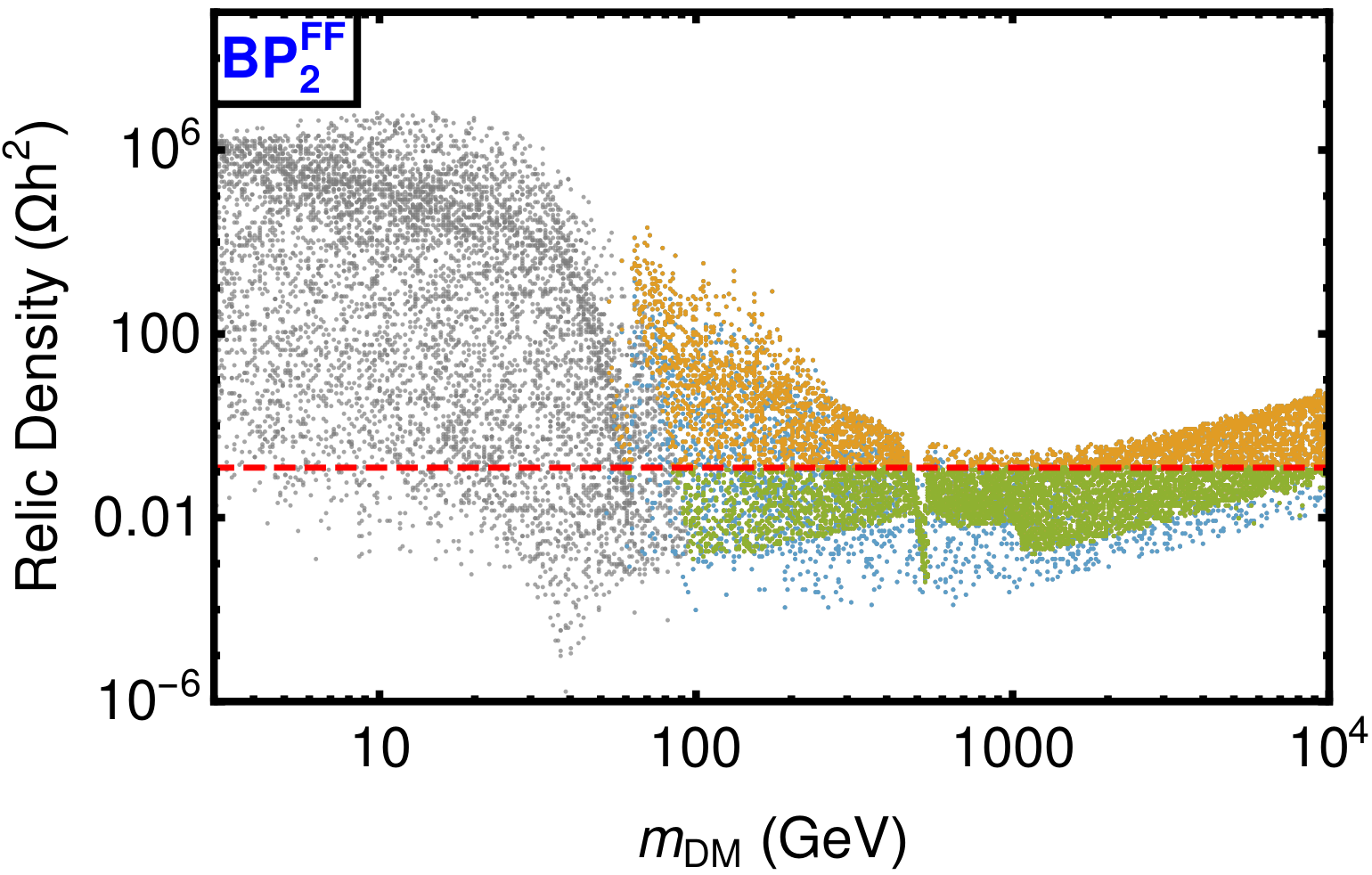}}
	\includegraphics[scale=0.4]{legend1}
	\caption{Relic density versus dark matter mass for the fermion-fermion co-annihilation case in Scenario-I (BP$_1^{\rm FF}$) and Scenario-II (BP$_2^{\rm FF}$).
          The red dashed line indicates the observed Planck value.}
	\label{fig:relicFF}
\end{figure}

\begin{figure}[h!]
	\scalebox{1.2}{\hspace{-1cm}
		\includegraphics[scale=0.13]{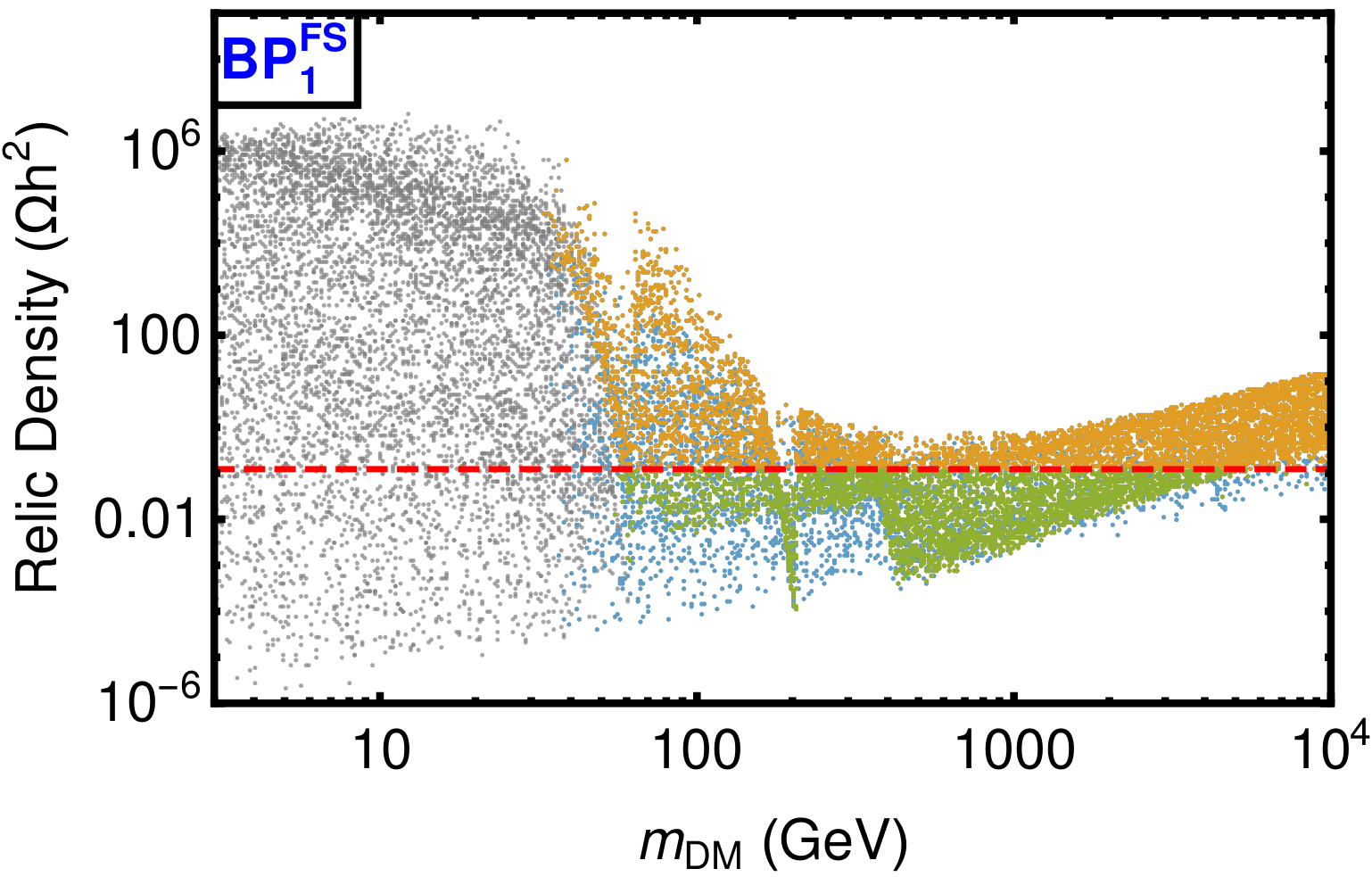}
		\hfill
		\includegraphics[scale=0.13]{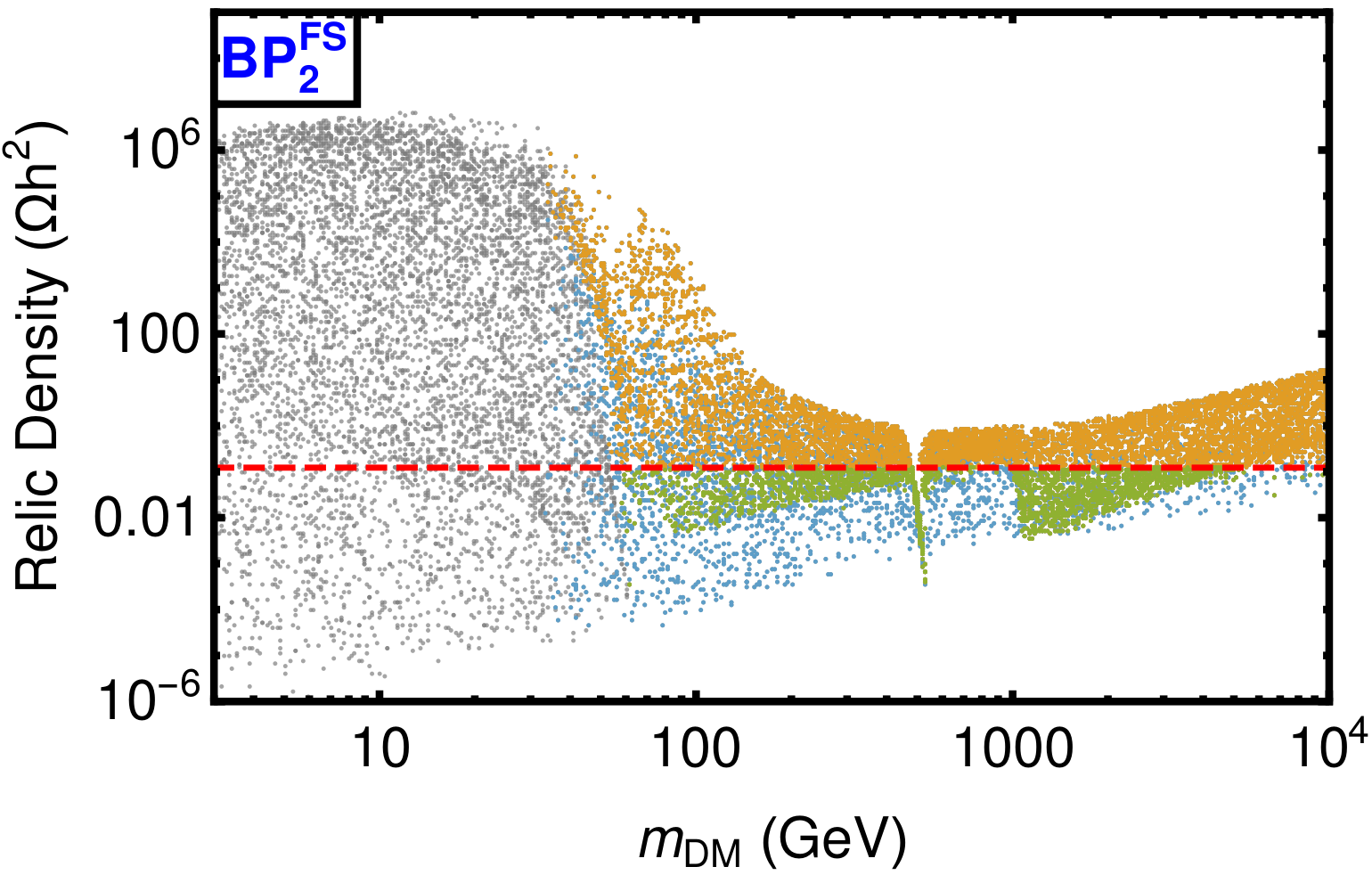}}
	\includegraphics[scale=0.4]{legend1}
	\caption{Relic density versus dark matter mass for the fermion-scalar co-annihilation case in Scenario-I (BP$_1^{\rm FS}$) and Scenario-II (BP$_2^{\rm FS}$).
          The red dashed line indicates the observed Planck value.}
	\label{fig:relicFS}
\end{figure}

Here we explore the singlet-like fermionic scotogenic dark matter relic density and the associated phenomenology, for masses ranging from 3 GeV to $10^4$ GeV. 
While Fig.~\ref{fig:relicnoco} describes the behaviour of the relic density for the no coannhilation case in Scenario-I and Scenario-II,
Fig.~\ref{fig:relicFF} and Fig.~\ref{fig:relicFS} illustrate the same when fermion-fermion and fermion-scalar co-annihilations are present. 
In these figures, the points excluded by the successive implementation of collider physics, cLFV and relic density constraints are indicated by gray, blue and orange colours, respectively.
In contrast, points obeying all of these restrictions are displayed in green colour. 

As discussed in Sec. \ref{sub:relic_a}, depending on the heavy Higgs boson mass $m_{H^0}^{}$ (whose values are $\sim400$ GeV and $\sim1100$ GeV for Scenario-I and Scenario-II, respectively)
one can clearly notice four different dips in relic density in all of the six plots from Fig.~\ref{fig:relicnoco} to Fig.~\ref{fig:relicFS}. 
Although the qualitative features of the rsults remain similar, scenario-II appears somewhat more constrained than the Scenario-I concerning the allowed regions given by the green points.\\[-2mm]

$\bullet$ {\bf No Co-annihilation} \\[-2mm]

In the no co-annihilation case, as indicated by BP$_1$ and BP$_2$ in \autoref{tab:BP12}, we vary both \dsf and \depf in the range of 100 GeV to 500 GeV, keeping \deip fixed \footnote{Choosing such \depf values always keeps the mass of $\eta^+$ above 100 GeV. Fixing \deip at 1000 GeV$^2$ then corresponds to a mass mass difference of less than 5 GeV between the charged and
    neutral components of $\eta$.} at 1000 GeV$^2$.  
This makes all $\mathcal Z_2$-odd particles very heavy compared to $\chi_1^0$, and hence the annihilation of singlet-like dark matter alone determines the freeze-out time and the relic density. 
As a result, all of the points in this case trivially satisfy the collider bounds mentioned in Sec. \ref{sub:Collider}. 
However, $\lambda_5$ and Im$(\omega)$ (which control $Y_F$ and $Y_\Sigma$) play a key role in determining the branching fractions for cLFV processes
as well as the relic fermioninc dark matter density in the region $m_{DM}<m_{H^0}^{}$. This can be seen from Fig.~\ref{fig:LFV} and Fig.~\ref{fig:relic2}. 
Therefore, the constraints on cLFV processes restrict $\lambda_5$ and Im$(\omega)$ in a severe way, ruling out most of the parameter-points below $m_{H^0}^{}$.
Indeed, in the left and right panels of Fig.~\ref{fig:relicnoco} one can see a cluster of blue points below the dark matter mass values of 400 GeV and 1100 GeV, respectively.  
Moreover, among the remaining points, a large portion gets ruled out by the relic density observed by Planck (marked by red dashed line). These are indicated by the orange colour. 
Thus, apart from the narrow regions near the scalar resonances, we mostly get allowed points (green) in the dark matter mass region beyond $m_{H^0}^{}$.  
Our results also indicate that it is not possible to get allowed points much beyond $10^4$ GeV. \\[-2mm]

 $\bullet$ {\bf Fermion-Fermion Co-annihilation} \\  
For the case of fermion-fermion co-annihilation, denoted by BP$_1^{\rm FF}$ and BP$_2^{\rm FF}$ in \autoref{tab:BP12}, we vary \dsf in the range of 1 GeV to 50 GeV.
This makes the masses of $\chi_{1,2}^0$ and $\Sigma^+$ close to each other, while the $\mathcal{Z}_2$-odd scalars ($\eta$) are much heavier.  
In contrast to the no co-annihilation case, here the lower mass region $m_{\rm DM}< m_{h^0}/2$ becomes entirely ruled out by the collider constraints only, see Fig.~\ref{fig:relicFF}.
These points are mainly eliminated by the constraint on $M_{\Sigma}$ ($m_{\Sigma^+})$ and the $h^0$ and $Z^0$ boson decay widths. 
Above $m_{h^0}/2$ dark matter mass values the restrictions on cLFV processes and relic density also play an important role. 
Nevertheless, new fermion-fermion dark matter co-annihilation channels open up, as shown in Fig.~\ref{fig:ffco}.   
This helps keeping the cLFV processes (or equivalently $Y_F$ and $Y_\Sigma$) under control.
Thus, in addition to the resonant region around $m_{h^0}/2$, we start getting allowed green points above 80 GeV dark matter mass in both Scenario-I and Scenario-II. 
As in the no co-annihilation case, we do not find allowed points much above $10^4$ GeV dark matter mass.\\[-2mm]

$\bullet$ {\bf Fermion-Scalar Co-annihilation} \\ 
The case of fermion-scalar co-annihilation is represented by BP$_1^{\rm FS}$ and BP$_2^{\rm FS}$ in \autoref{tab:BP12}. 
Now \depf is varied within the range from 1 GeV to 30 GeV with \deip varying from 1 GeV$^2$ to 1000 GeV$^2$, keeping very large $\chi_2^0$ masses. 
In this case, the light dark matter region of $m_{\rm DM}< m_{h^0}/2$ is disfavoured because of collider constraints on the $h^0$ and $Z^0$ boson decay widths. 
Above $m_{h^0}/2$, bounds from cLFV and relic density also dismiss plenty of points. 
Yet, in both panels of Fig.~\ref{fig:relicFS}, one can see a broad allowed region of green points above 62.5 GeV. 
The presence of various processes depicted in Fig.~\ref{fig:fsco} assists in the lessening of the relic density in this case.
It is interesting to mention that, unlike the earlier two cases, here dark matter masses above 5000 GeV become ruled out.

\vspace*{-5mm}
%%%%%%%%%%%%%%%%%%%%%%%%%%%%%%%%%%%%
\subsection{Direct Detection}  
\label{sec:direct-detection}
\vspace*{-3mm}

\begin{figure}[h!]
	\centering
	\hspace*{-1.0cm}
	\includegraphics[scale=0.47]{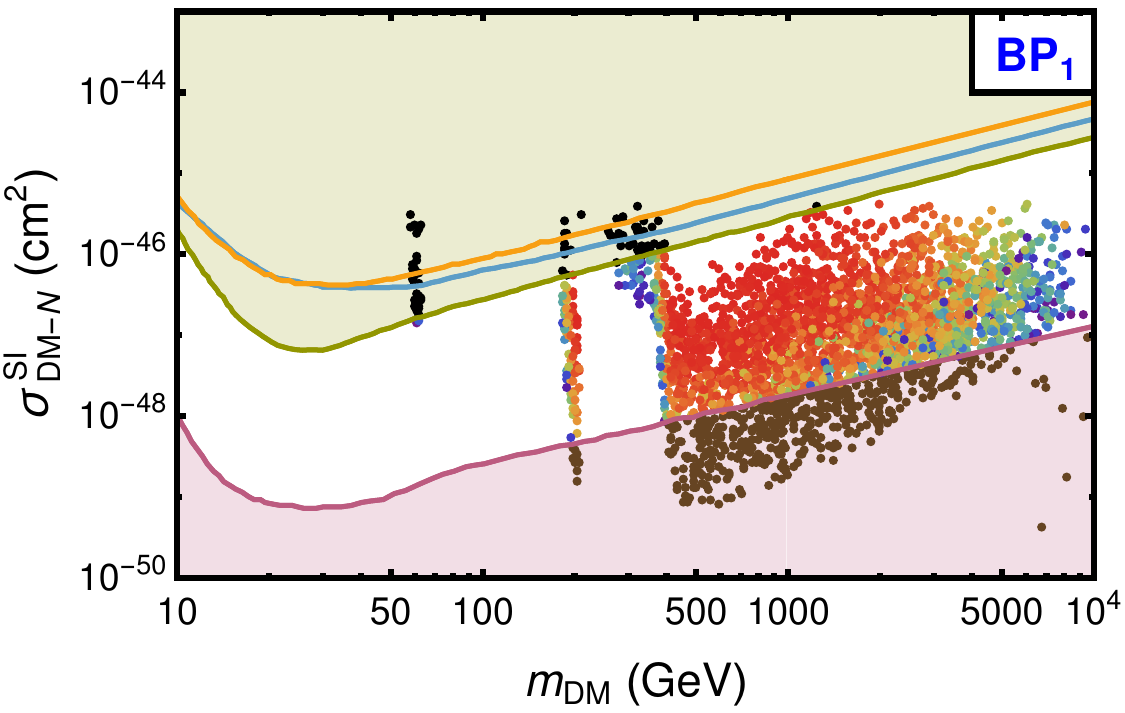}
	\hfil
	\includegraphics[scale=0.47]{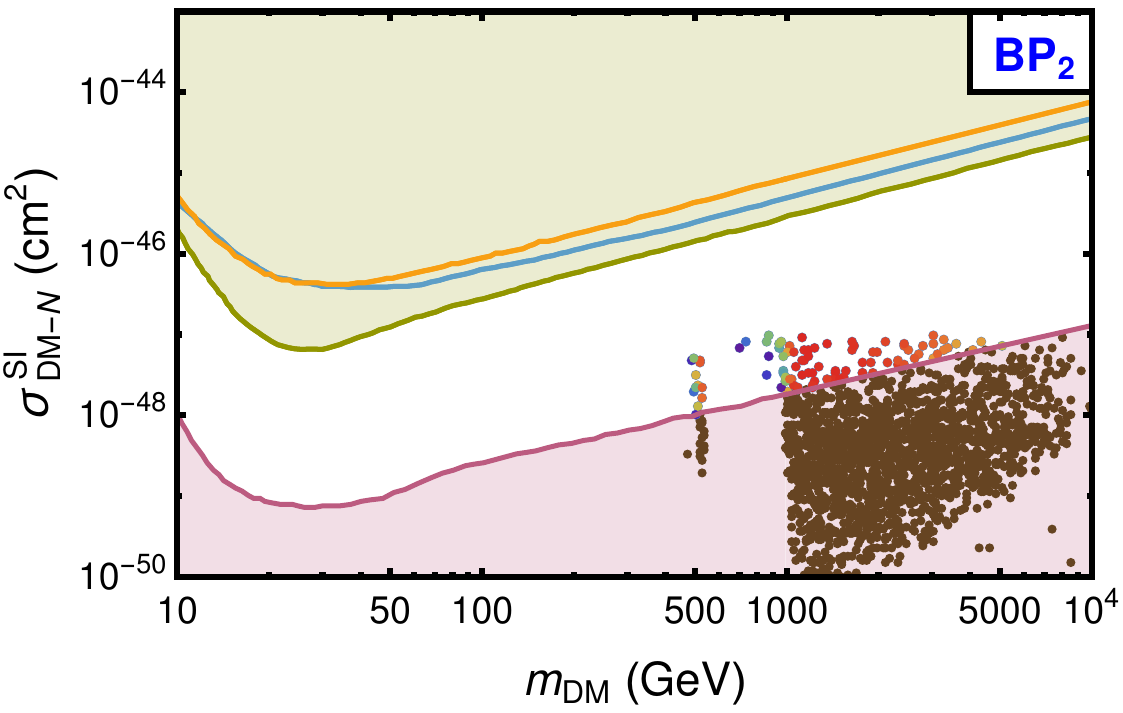}
	\text{\red \large \textbf{No Coannihilation}}
	
	\vspace*{0.5cm}\hspace*{-1.4cm}
	\includegraphics[scale=0.47]{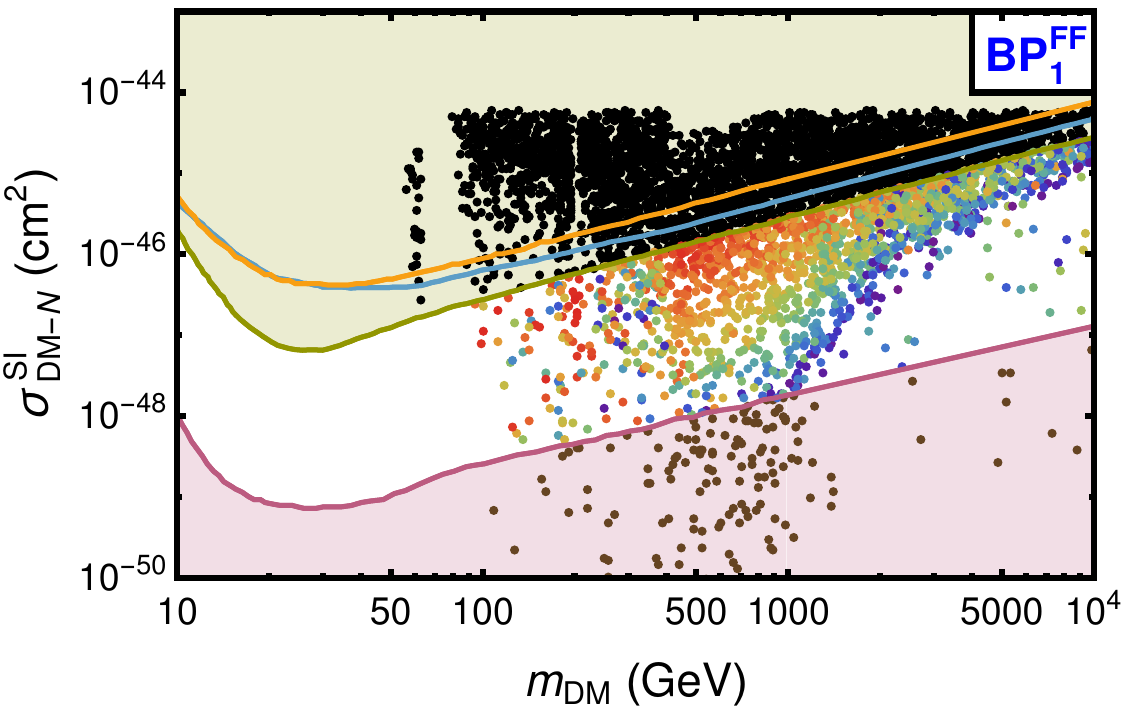}
	\hfil
	\includegraphics[scale=0.47]{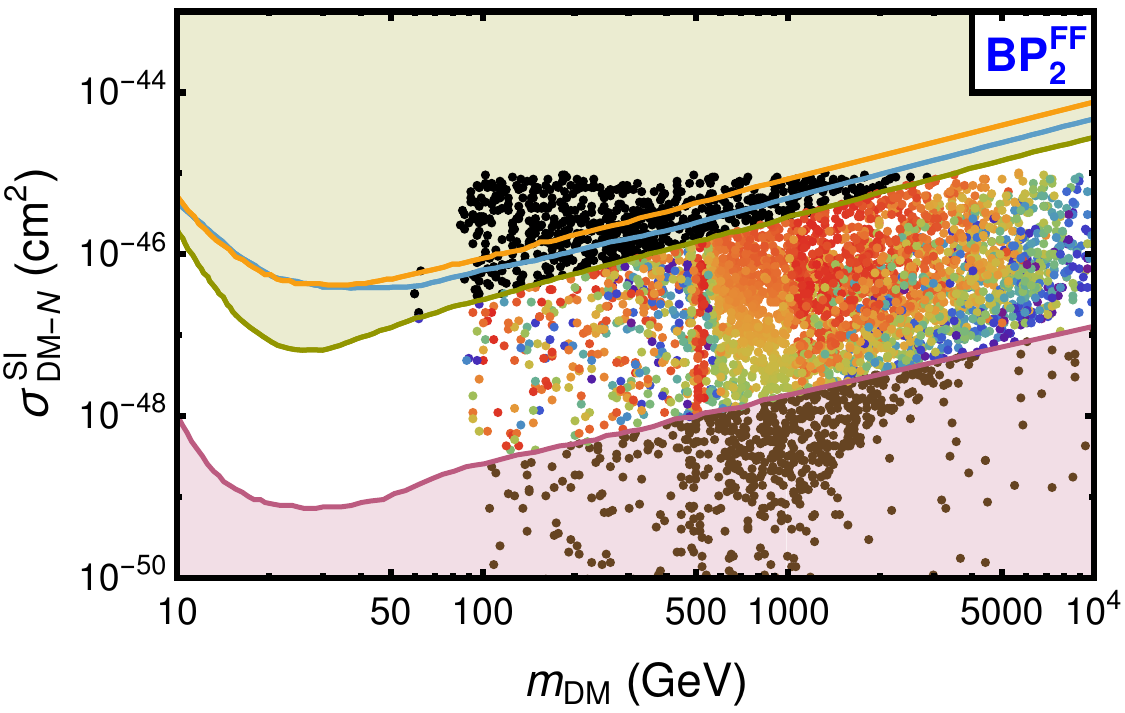}
	\text{\red \large \textbf{Fermion-Fermion Coannihilation}}
	
	\vspace*{0.5cm}\hspace*{-1.4cm}
	\includegraphics[scale=0.47]{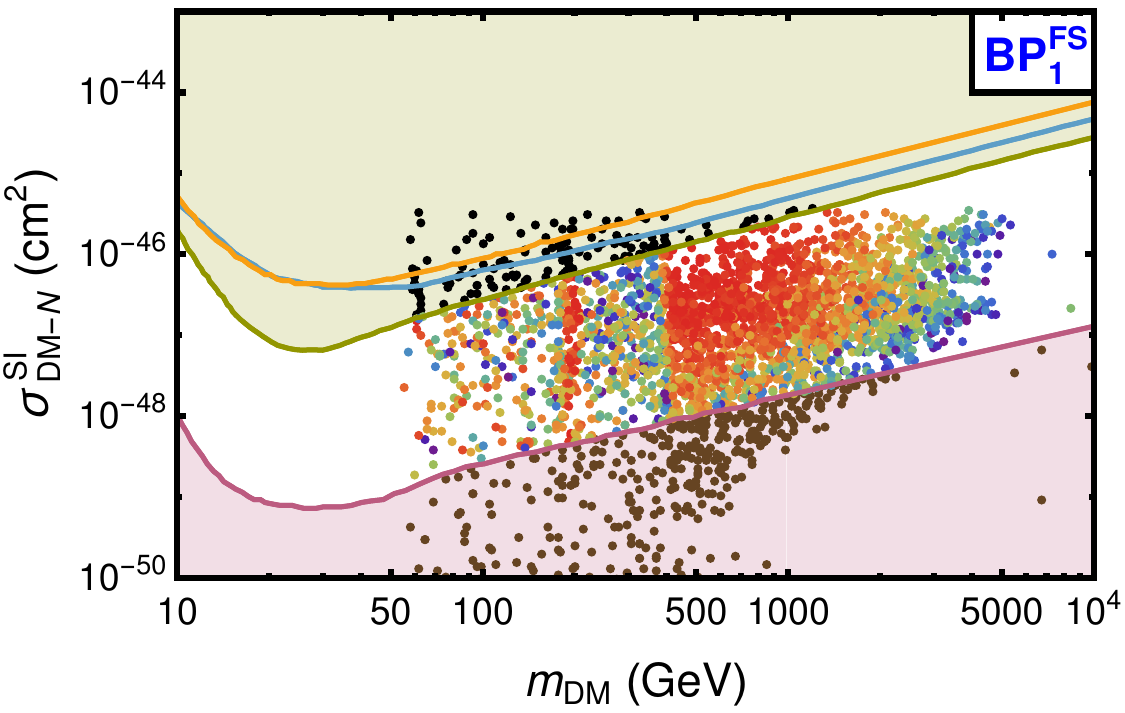}
	\hfil
	\includegraphics[scale=0.47]{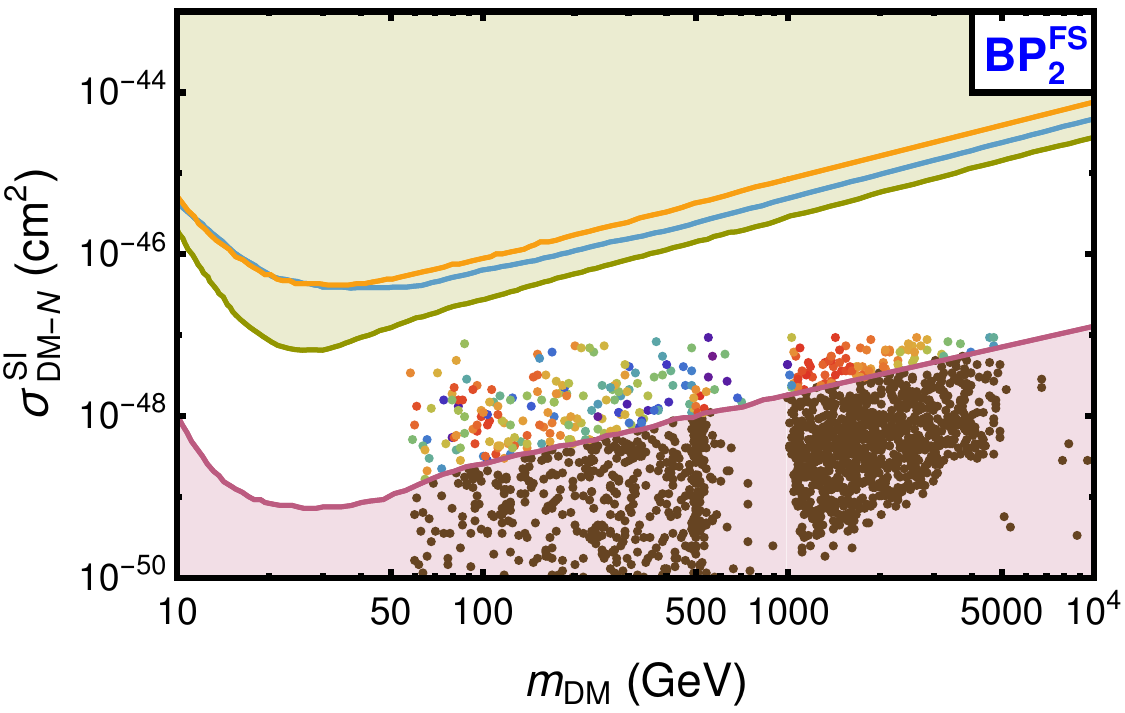}
	\text{\red \large \textbf{Fermion-Scalar Coannihilation}}

	\vspace*{0.2cm}
	\scalebox{1.1}{\hspace{-0.6cm}\includegraphics[height=1.4cm,width=4.5cm]{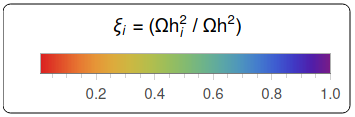}\hfill
		\raisebox{-0.2cm}{\includegraphics[height=1.8cm,width=13.2cm]{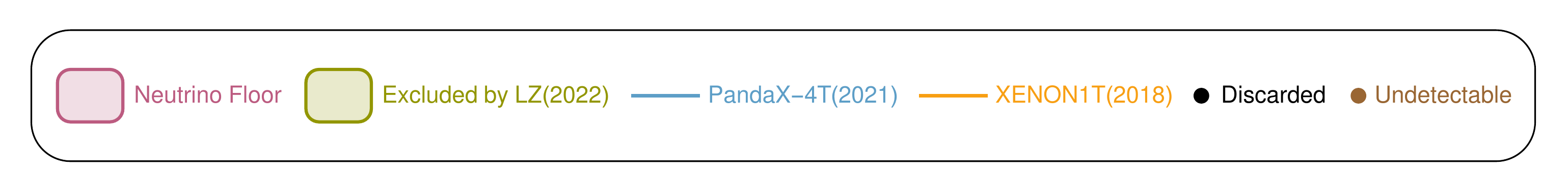}}}
	\caption{Spin-independent dark matter-nucleon scattering cross section (in cm$^2$) for Scenario-I and Scenario-II.
		Parameter regions with points satisfying $1\%<\xi_i<100\%$ where $\xi_i=(\Omega h_i^2/\Omega h^2)$ are indicated in colour.}
	\label{fig:DD_sc}
\end{figure}

Now we discuss the direct detection prospects of fermionic scotogenic dark matter in various experiments such as LZ, Xenon-nT, PandaX.
In Fig.~\ref{fig:DD_sc} we give direct detection results for all cases of no co-annhilation, fermion-fermion and fermion-scalar co-annihilation, both in Scenario-I and Scenario-II.
The purple region indicates the neutrino floor arising from coherent scattering of neutrinos,
whereas the greenish area represents the region ruled out by the LZ direct detection experiment, that currently provides the strongest restriction. 
In our study, we collect the green points from Fig.~\ref{fig:relicnoco}, Fig.~\ref{fig:relicFF} and Fig.~\ref{fig:relicFS} giving the observed relic density
or providing at least 1\% of the entire observed relic density.
The resulting spin-independent dark matter-neucleon scattering cross-section is presented in Fig.~\ref{fig:DD_sc}. 
The points ruled out by direct detection experiments are coloured in black, while the undetectable points below the neutrino floor are displayed in brown.
All the other points are viable for fermionic scotogenic dark matter in our triplet-singlet model.

We start in the left column in Fig.~\ref{fig:DD_sc} by depicting the direct detection prospects for Scenario-I with $v_\Omega = 4~$GeV,  
while Scenario-II with $v_\Omega = 1.5~$GeV is examined in the right column in Fig.~\ref{fig:DD_sc}. 
One sees that in all cases the clusters of points follow a similar pattern in both scenarios.
We also notice that the maximum dark matter-nucleon scattering cross section decreases for all the three cases in Scenario-II, when compared to that in the Scenario-I.
Indeed, according to  Eq. \eqref{eq:tht_mch} and  Eq. \eqref{eq:beta}, a lower $v_\Omega$ pushes both scalar and fermion mixing angles ($\beta$ and $\theta$, respectively) to smaller values.
This reduces the direct detection cross section to a great extent, see  Eq. \eqref{eq:DD_cr}.

One sees that dark matter detection is possible in Scenario-I (left column in Fig.~\ref{fig:DD_sc}) in all cases of no co-annihilation (BP$_1$),
fermion-fermion co-annihilation (BP$_1^{\rm FF}$) as well as fermion-scalar co-annihilation (BP$_1^{\rm FS}$). 
Nevertheless we find the last case of fermion-scalar co-annihilation to be most promising.
Indeed, one obtains a broad well-distributed set of points within the dark matter detectability region (white) all the way from 62.5 GeV $(m_{h^0}/2)$  upto 5 TeV.
In contrast, for the no co-annihilation case, most of the detectable points lie above 400 GeV (i.e. $m_{H^0}$).  
On the other hand, in the case of fermion-fermion co-annihilation, most of the points get ruled out by direct detection experiments.  
Due to the smaller value of \dsf characteristic of this case, the dark fermion fermionic mixing angle $\theta$ increases.
This leads to a substantial enhancement in the direct detection cross section $\sigma_{\rm DM-N}^{\rm SI}$. 
Although one gets detectable points for dark matter masses between 100 GeV to 10 TeV, they are not very well-distributed inside the white region. 
In contrast, in the fermion-scalar co-annihilation case, the larger \dsf pushes $\sigma_{\rm DM-N}^{\rm SI}$ below the direct detection upper limits. 
Moreover, co-annihilation effects help to keep the relic density small.
This makes fermion-scalar co-annihilation into a viable option for fermionic scotogenic dark matter in the case of higher $v_\Omega$. 

We now turn to the possibility of having lower value of $v_\Omega$, as in Scenario-II, i.e. 1.5 GeV, shown in the right column in Fig.~\ref{fig:DD_sc}. 
Clearly the fermion-fermion co-annihilation option appears to be the best option for having detectable fermionic scotogenic dark matter in Scenario-II. 
The combined effect of having large $ \Delta m_{\Sigma F}$ and small $v_\Omega$ in the case of no co-annihilation or fermion-scalar co-annihilation leads to smaller cross sections
$\sigma_{\rm DM-N}^{\rm SI}$ in Scenario-II, that go below the neutrino floor, making a big chunk of the parameter space undetectable. 
In contrast, the smaller value of $ \Delta m_{\Sigma F}$ enhances $\sigma_{\rm DM-N}^{\rm SI}$ in the fermion-fermion co-annihilation.
This ensures good detectability prospects for the dark matter masses above 100 GeV in this case. 
Thus for lower $v_\Omega$ fermion-fermion co-annihilating fermionic scotogenic dark matter becomes a perfect candidate to be explored.

 \vspace*{-5mm}
\section{Conclusions and outlook} 
\label{sec:conclusions}
\vspace*{-3mm}

We have explored in a dedicated manner the possibility that neutrino masses arise from the exchange of dark matter states, as in Fig.~\ref{fig:nu_mas}.
The same $\mathcal Z_2$ symmetry that makes the neutrino masses calculable also stabilizes the dark matter.
We examined the rich phenomenological profile of scotogenic fermionic dark matter within the singlet-triplet model proposed in~\cite{Hirsch:2013ola}.
In this reference model we stressed the possibility of having a singlet-like fermion $\chi_1^0$ as dark matter candidate~\footnote{Viable WIMP triplet-like dark matter requires masses $\sim 3$ TeV, in order to produce the observed relic density.}.
This updates and substantially extends the original work in~\cite{Hirsch:2013ola} and also complements the work in Refs.~\cite{Restrepo:2019ilz,Choubey:2017yyn}
and~\cite{Diaz:2016udz,Avila:2019hhv}, the latter dealing with the phenomenology of scotogenic scalar dark matter. 
We determined the allowed parameter region in this model consistent with the imposition of various theoretical and experimental restrictions. 
These include consistency conditions for the scalar potential, electroweak precision observables, constraints from neutrino oscillations,
collider experiments, as well as cosmological relic density of dark matter.
Charged lepton flavour violation proceeds vis the exchange of dark states, see Figs.~\ref{fig:mutoe}.
Taking into account all of these we have determined the prospects for cLFV searches, see Fig.~\ref{fig:LFV}.

In order to get a fuller picture of the phenomenological profile of fermionic dark matter in our scenario we combined our relic density results with our direct detection analysis.
The cosmological relic density is set by the processes indicated in Fig.~\ref{fig:ann-DM}.
Our relic density results are given in Figs.~\ref{fig:relic0}, \ref{fig:relic1}, \ref{fig:relic2} and \ref{fig:relic3}, under various specific parameter assumptions. 
To obtain a more global picture we have performed a dedicated numerical scanning procedure.
To make it efficient we have made a number of reasonable assumptions so as to reduce the number of independent new physics parameters.
Our results are summarized in Figs.~\ref{fig:relicnoco}, \ref{fig:relicFF} and \ref{fig:relicFS}, see detailed discussion in the text.
For example, we noted that low-mass dark matter ($\lesssim m_{h^0}/2$) with viable relic density is in conflict with cLFV and/or collider limits. 
Our analysis indicates that, in the absence of co-annihilation, Figs.~\ref{fig:ffco} and \ref{fig:fsco}, allowed fermionic dark matter masses typically require $m_{\rm DM} \gtrsim m_{H^0}$.
In fact, dark matter with masses upto 100 GeV will be viable only in the presence of scalar-fermion co-annihilation, see lower left and right panel in Fig.~\ref{fig:DD_sc}.

In contrast to the original scotogenic constructions~\cite{Ma:2006km,Tao:1996vb} 
dark matter detection in nuclear recoil experiments proceeds at the tree level~\cite{Hirsch:2013ola}, through the diagram in Fig.~\ref{fig:DD}.
Our present study indicates that the generalized fermionic scotogenic dark matter scenario leads to promising direct detection prospects, as summarized in Fig.~\ref{fig:DD_sc}.

In summary, we conclude that fermionic scotogenic dark matter brings in novel features with respect to \emph{vanilla} fermionic dark matter
in supersymmetry~\cite{Jungman:1995df,Bertone:2016nfn,Arcadi:2017kky}. 
Indeed, we note that our scenario does mimic neutralino dark matter without having an underlying supersymmetric framework.
For example, our singlet-like fermionic dark matter is very much analogous to the ``bino-like'' fermionic supersymmetric WIMP. 
However, for bino-like fermionic dark matter, even if viable relic densities are obtained through co-annihilation~\cite{Ellis:1998kh,Profumo:2004wk,Boehm:1999bj},
direct detection cross sections typically remain below the neutrino floor~\cite{Baer:2005jq,Duan:2018rls}, in contrast with results we have obtained in Fig.~\ref{fig:DD_sc}.

Moreover, fermionic scotogenic dark matter is accompanied by the presence of cLFV phenomena.
Although absent in the case of supersymmetric dark matter, these constitute an intrinsic feature in the scotogenic dark matter construction.
As a final comment we stress that the ``missing-partner'' nature of the radiative seesaw mechanism mediated by the dark fermions,
  see \autoref{tab:part}, implies the presence of a massless neutrino, with a very clear implication concerning neutrinoless double beta decay.
  Indeed, the amplitude for latter has a lower bound~\cite{Avila:2019hhv,Reig:2018ztc,Barreiros:2018bju} which could make it detectable in the next round of
  experiments, for the case of inverse neutrino mass-ordering. Together with cLFV phenomena, the possible observation of \znbb would provide another distinctive feature
  of scotogenic fermionic dark matter with respect to vanilla supersymmetric dark matter.
All in all, our generalized scotogenic framework provides a richer picture of particle dark matter than its supersymmetric counterpart.
Upcoming experiments will help testing the two approaches in a quantitative manner.

\vspace*{-5mm}
%%%%%%%%%%%%%%%%%%%%%%%%
\begin{acknowledgments}
%%%%%%%%%%%%%%%%%%%%%%%%
\vspace*{-3mm}

  We thank M. Hirsch and S. Heinemeyer for useful discussions.
  This work was supported by the Spanish grants PID2020-113775GB-I00~(AEI/10.13039/501100011033) and Prometeo CIPROM/2021/054 (Generalitat Valenciana)
  and by                                        PID2020-114473GB-I00~(MCIN/AEI/10.13039/501100011033) and Prometeo/2021/071 (Generalitat Valenciana). 
  S.S. thanks SERB, DST, Govt. of India for a SIRE grant with number SIR/2022/000432, supporting an academic visit to AHEP group at IFIC.
  We thank Avelino Vicente for discussions.
  %%%%%%%%%%%%%%%%%%%%%% 
\end{acknowledgments}
%%%%%%%%%%%%%%%%%%%%%%%%

\appendix 
\label{Appendix}
\vspace*{-5mm}
\section{Changes in the decay widths of the Higgs, $Z$ and $W$}
\vspace*{-3mm}

In our Singlet-Triplet Scotogenic Model the $Z$,~$W$ and Higgs bosons of the Standard Model may have new contributions to their decays associated to the new particles present, as follows.
\label{sec:width}
\begin{align}
&\Gamma_{Z\to\eta_I \eta_R}=\frac{e^2 \,m_Z}{48\pi\sin^22\theta_w}\,\lambda^{3/2}\Big(1,\frac{m_{\eta_R}^2}{m_Z^2},\frac{m_{\eta_I}^2}{m_Z^2}\Big),\\
&\Gamma_{Z\to\eta^+ \eta^-}=\frac{e^2 \,m_Z}{48\pi\tan^22\theta_w}\,\Big(1-\frac{4m_{\eta^\pm}^2}{m_Z^2}\Big)^{3/2},\\
&\Gamma_{Z\to\chi^+ \chi^-}=\frac{e^2 \,m_Z}{12\pi}\,\Big(1-\frac{4m_{\chi^\pm}^2}{m_Z^2}\Big)^{1/2}\,\Big(1+\frac{2m_{\chi^\pm}^2}{m_Z^2}\Big),\\
&\Gamma_{W^\pm\to\chi^\pm \chi_j}=\frac{e^2\, m_W \,R_{\chi_{j2}}^2}{24\pi\sin^2\theta_w}\Big[m_W^2-(m_{\chi^\pm}-m_{\chi_j})^2\Big] \Big[2m_W^2+(m_{\chi^\pm}+m_{\chi_j})^2\Big]\lambda^{1/2}\Big(1,\frac{m_{\chi^\pm}^2}{m_W^2},\frac{m_{\chi_j}^2}{m_W^2}\Big),\\
&\Gamma_{W^\pm\to\eta^\pm \eta_k}=\frac{e^2\, m_W}{192\pi\,\sin^2\theta_w}\lambda^{3/2}\Big(1,\frac{m_{\eta^\pm}^2}{m_W^2},\frac{m_{\eta_k}^2}{m_W^2}\Big),\\
&\Gamma_{h\to\eta^+ \eta^-}=\frac{1}{16\pi\,m_h}\Big(1-\frac{4m_{\eta^\pm}^2}{m_h^2}\Big)^{1/2}\Big[\lambda_3 v_\phi R_{h_{11}}+\Big(\frac{\mu_2}{\sqrt 2}+\lambda_{\eta}^\Omega\, v_\Omega\Big)R_{h_{12}}\Big]^2,\\
&\Gamma_{h\to\eta_k \eta_k}=\frac{1}{32\pi\,m_h}\Big(1-\frac{4m_{\eta_k}^2}{m_h^2}\Big)^{1/2}\Big[(\lambda_3+\lambda_4+c_k\lambda_5) v_\phi R_{h_{11}}-\Big(\frac{\mu_2}{\sqrt 2}-\lambda_{\eta}^\Omega\, v_\Omega\Big)R_{h_{12}}\Big]^2,\\
&\Gamma_{h\to\chi_i^{} \chi_j^{}}=\frac{Y_\Omega^2\,R_{h_{12}}^2 d_{ij}}{16\pi\,m_h}\lambda^{1/2}\Big(1,\frac{m_{\chi_i}^2}{m_h^2},\frac{m_{\chi_j}^2}{m_h^2}\Big)\Big[m_h^2-(m_{\chi_i^{}}+m_{\chi_j^{}})^2\Big](R_{h_{i1}}R_{h_{j2}}+R_{h_{i2}}R_{h_{j1}})^2,
\end{align}
where, $(i,j)\in\{1,2\}$, $k\in\{R,I\}$ corresponding to real and imaginary parts, $c_k$ is $\pm1$ for $\eta_{R,I}$ respectively and $d_{ij}$ equals to one or two depending on whether $i=j$ or $i\neq j$ respectively.
Here $R_{\chi_{\alpha\beta}^{}}$ and $R_{h_{\alpha\beta}^{}}$ denote the $(\alpha,\beta)$ elements of the rotation matrices diagonalizing the neutral fermionic mass matrix $(\mathcal M_\chi)$
and the squared neutral scalar mass matrix $(\mathcal M_h^2)$ respectively.
In the above expressions, one must impose the condition that mass of the decaying particle is greater than the sum of masses for the final state particles. 

\vspace*{-5mm}
\section{Co-annihilation}
\label{sec:co}
\vspace*{-3mm}

\begin{figure}[h!]
	\includegraphics[scale=0.2]{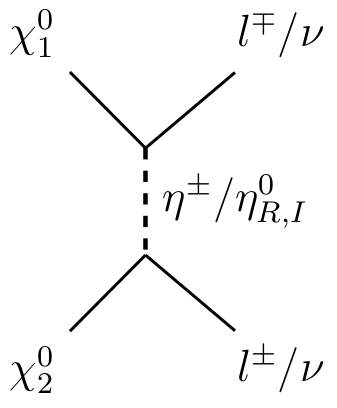}
	\includegraphics[scale=0.2]{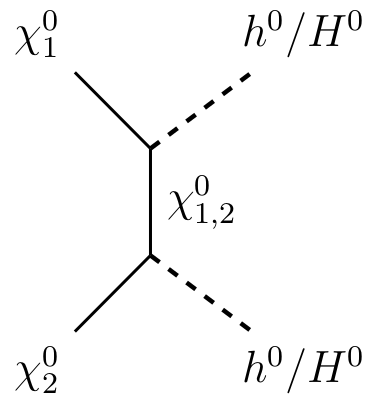}
	\includegraphics[scale=0.2]{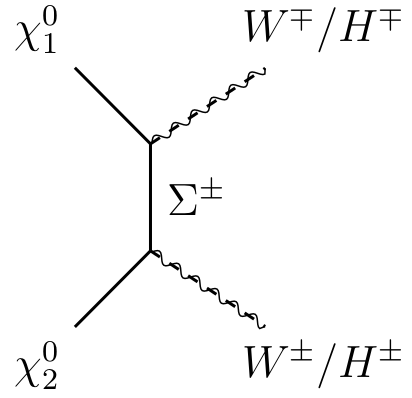}
	\includegraphics[scale=0.2]{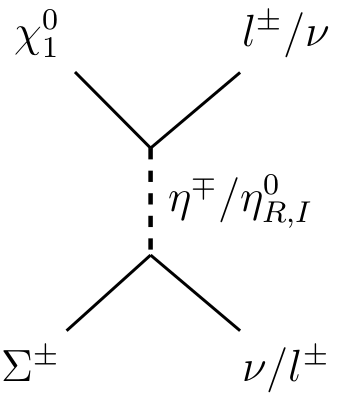}
	\includegraphics[scale=0.2]{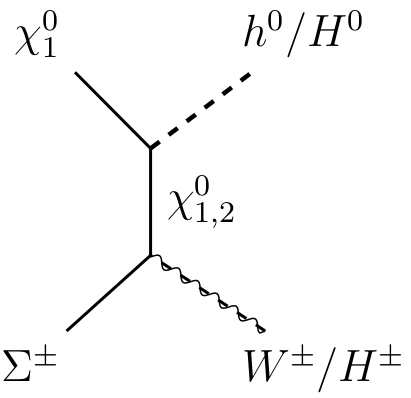}
	\includegraphics[scale=0.2]{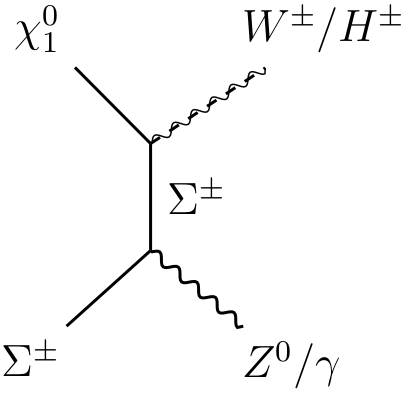}
	\vspace*{5mm}
	
	\includegraphics[scale=0.2]{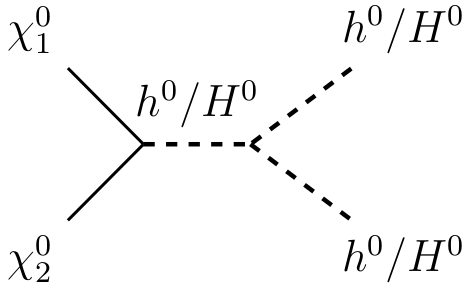}
	\includegraphics[scale=0.2]{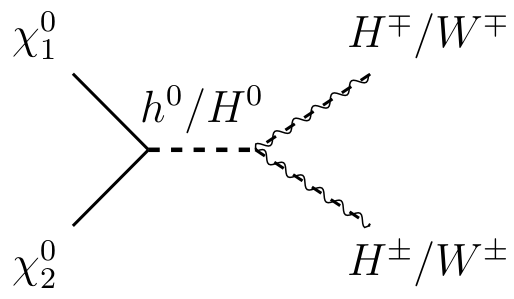}
	\includegraphics[scale=0.2]{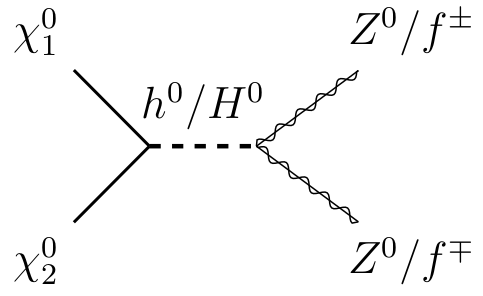}
	\includegraphics[scale=0.2]{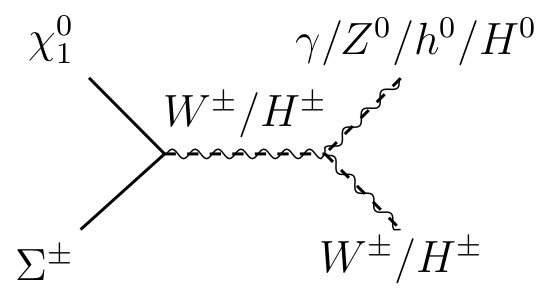}
	\includegraphics[scale=0.19]{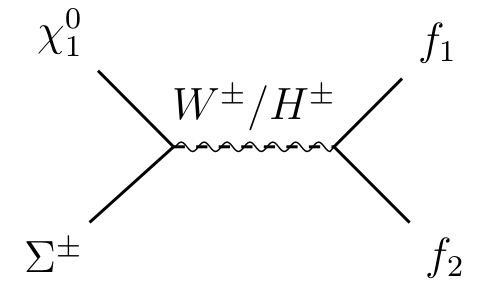}
	\caption{Feynman diagrams for processes involved fermion-fermion co-annihilation.}
	\label{fig:ffco}
\end{figure}

\begin{figure}[h!]
	\scalebox{1.1}{
	\hspace{-0.5cm}
	\includegraphics[scale=0.15]{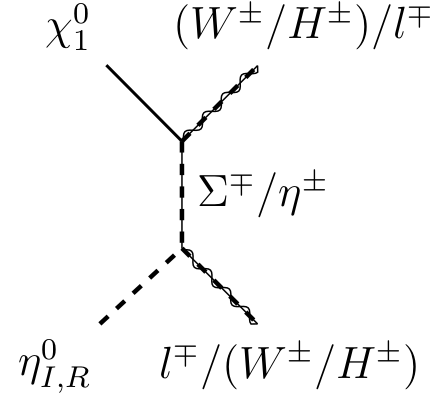}\hspace{-1.5mm}
	\includegraphics[scale=0.15]{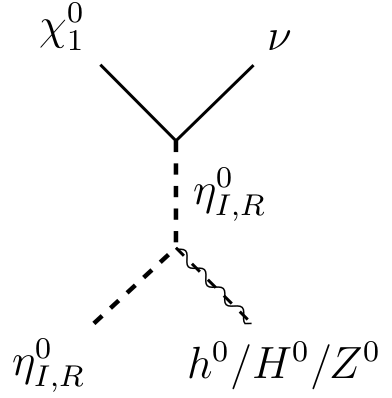}\hspace{-1.5mm}
	\includegraphics[scale=0.15]{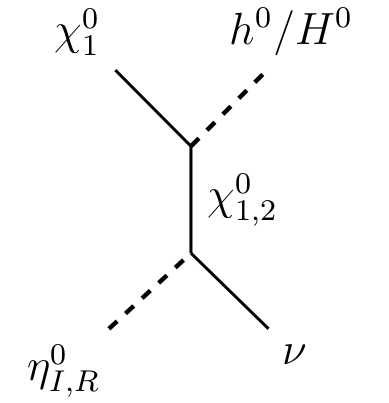}\hspace{-2.5mm}
	\includegraphics[scale=0.15]{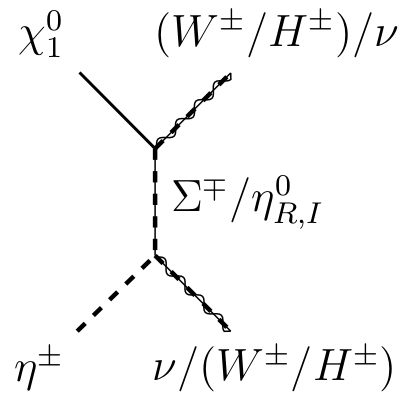}
	\raisebox{4.0mm}[0pt][0pt]{\hspace{-2.5mm}
	\includegraphics[scale=0.15]{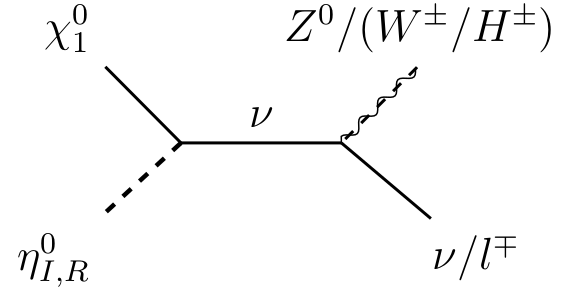}\hspace{-1.0mm}
	\includegraphics[scale=0.15]{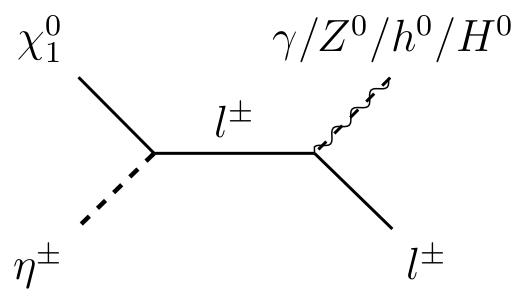}\hspace{-1.0mm}
	\includegraphics[scale=0.15]{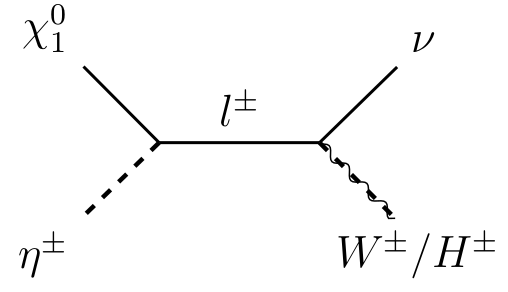}}}
	
	\caption{Feynman diagrams for processes involved fermion-scalar co-annihilation.}\label{fig:fsco}
\end{figure}

Below we give the Feynman diagrams involved in dark matter co-annihilation. In addition to Fig.~\ref{fig:ffco} depicting fermion-fermion co-annihilation,
we give also Fig.~\ref{fig:fsco} which corresponds to fermion-scalar co-annihilation processes.

\vspace*{-8mm}
%\newpage
\bibliographystyle{utphys}
%\bibliography{bibl} 
\bibliography{bibliography} 

\end{document}